\documentclass[pra,twocolumn,showpacs]{revtex4-1}
\usepackage{amsmath,amscd,amsfonts,amssymb,color}
\usepackage{graphicx,amsfonts,epsf}
\usepackage{epstopdf}
\usepackage{hyperref}
\usepackage{ulem}

\newcommand{\ie}{\textit{i.e.~}}
\begin{document}
\title{Experimentally identifying the entanglement class of pure
tripartite states}
\author{Amandeep Singh}
\email{amandeepsingh@iisermohali.ac.in}
\affiliation{Department of Physical Sciences, Indian
Institute of Science Education \& 
Research Mohali, Sector 81 SAS Nagar, 
Manauli PO 140306 Punjab India.}
\author{Kavita Dorai}
\email{kavita@iisermohali.ac.in}
\affiliation{Department of Physical Sciences, Indian
Institute of Science Education \& 
Research Mohali, Sector 81 SAS Nagar, 
Manauli PO 140306 Punjab India.}
\author{Arvind}
\email{arvind@iisermohali.ac.in}
\affiliation{Department of Physical Sciences, Indian
Institute of Science Education \& 
Research Mohali, Sector 81 SAS Nagar, 
Manauli PO 140306 Punjab India.}
\begin{abstract}
We use concurrence as an entanglement measure and experimentally demonstrate
the entanglement classification of arbitrary three-qubit pure states on a
nuclear magnetic resonance (NMR) quantum information processor.  Computing the
concurrence experimentally under three different bipartitions, for an arbitrary
three qubit pure state, reveals the entanglement class of the state.  The
experiment involves measuring the expectation values of Pauli operators.  This
was achieved by mapping the  desired expectation values onto the local $z$
magnetization of a single qubit.  We tested the entanglement classification
protocol on twenty seven different generic states and successfully detected
their entanglement class.  Full quantum state tomography was performed to
construct experimental tomographs of each state and negativity was calculated
from them, to validate the experimental results.
\end{abstract} 
\pacs{03.67.Mn} 
\maketitle 
\section{Introduction} 
It is a well established fact that quantum entanglement is a key resource to
achieve computational speedup in quantum information processing (QIP)
tasks~\cite{horodecki-rmp-09}.  Entanglement characterization and detection is
of utmost importance for the physical realization of quantum information
processors~\cite{guhne-pr-09,li-ff-13}.  The presence of entanglement can be
confirmed using several methods such as quantum state
tomography~\cite{thew-pra-02}, witness operators~\cite{guhne-jmo-03,
arrazola-pra-12,jungnitsch-prl-11}, the density operator under partial
transposition~\cite{peres-prl-96,li-sr-17} and via the violation of Bell's
inequalities~\cite{peres-pra-97}.

Entangled states have been physically realized in superconducting phase
qubits~\citep{neeley-nature-10}, nitrogen-vacancy defect
centers~\citep{neumann-science-08}, nuclear spin qubits~\citep{dogra-pra-15},
quantum dots~\citep{gao-nature-12} and trapped-ion~\citep{mandel-nature-03}
quantum computing hardwares.  Entanglement creation and detection has been
demonstrated in
NMR~\cite{laflamme-rsta-98,peng-pra-10,rao-ijqi-12,dogra-pra-15-1,xin-pra-18}
and pseudo-bound entanglement was detected using a three-qubit
system~\citep{kampermann-pra-10}.  There are several measures to quantify and
detect the entanglement~\citep{guhne-pr-09, dagmar-jmp-02}. The entanglement of
formation was used as an entanglement quantifier in four trapped
ions~\citep{sackett-nature-00}, and concurrence~\citep{wootters-qic-01} was
measured in a single experiment on twin copies of the quantum state of
photons~\citep{walborn-nature-06}. Entanglement was also explored using witness
based detection protocols in NMR~\citep{filgueiras-qip-12} as well as in
quantum optics~\citep{bourennane-prl-04}.

The characterization and detection of multipartite
entanglement is a challenging task in terms of the required
experimental and computational
resources~\cite{dur-jpamg-01,altepeter-prl-05,spengler-pra-12,dai-prl-14}
It is hence important to design and experimentally implement
entanglement detection protocols which use fewer resources.
Three-qubit states have been
classified into six inequivalent classes~\citep{dur-pra-00}
under stochastic local operation and classical communication
(SLOCC)~\citep{bennett-pra-00}  and  several
protocols~\cite{chi-pra-2010,zhao-pra-13,akbari1,akbari2}
have been proposed to ascertain the class of a given three
state.

In the present study, we experimentally characterize the
entanglement class of arbitrary three-qubit pure states.
Towards achieving our goal we utilize 
a concurrence-based~\citep{wootters-prl-98, wootters-qic-01}
entanglement classification protocol proposed by Zhao et.
al~\citep{zhao-pra-13}.  The advantage of this protocol is
that it can be realized for any three-qubit pure state, as
compared to previous proposals which are limited to the
class of three-qubit generic
states~\cite{adhikari-arxiv-17,singh-arxiv-18}.  The
experimental implementation relies on efficiently
determining the expectation values of desired Pauli
operators and to achieve this we 
used a previously designed scheme which maps the
expectation values onto the local $z$ magnetization of
a single qubit~\citep{singh-pra-16}.
A total of twenty
seven states were prepared to experimentally implement the
protocol: seven representative states belonging to the six
SLOCC inequivalent classes and twenty randomly generated
states, with state fidelities ranging between 88\% to 99\%.
The protocol successfully identified the entanglement class
of all the seven representative states (namely, GHZ, W, $\rm
W\overline{W}$, three bi-separable states and a separable
state) within the experimental error limits.  Further, the
randomly generated three-qubit states were also classified
successfully as belonging to either the GHZ, the W, the
bi-separable or the separable class of states.  Full quantum
state tomography~\citep{leskowitz-pra-04} 
was performed and the
entanglement measure negativity~\citep{weinstein-pra-10,vidal-pra-02}
was computed from the experimentally reconstructed
state, to validate the experimental
results.

The paper is organized as follows: Section~\ref{Theory} 
outlines the theoretical
framework for three-qubit entanglement
classification where we  describe the classification
protocol. In Section~\ref{NMR Implementation}
the NMR implementation of the protocol and our main
results are  described.
Concluding remarks are contained in 
Section~\ref{remarks}.
\section{Three Qubit Pure State Entanglement Classification}
\label{Theory}
Consider a three-qubit pure state $ \vert \Psi \rangle $. 
The state is fully
separable if one can write $ \vert \Psi \rangle = \vert \psi_1 \rangle \otimes
\vert \psi_2 \rangle \otimes \vert \psi_3 \rangle $. In case $ \vert \Psi
\rangle $ is biseparable under bipartition $ 1 \vert 23 $, then it is always
possible to write $ \vert \Psi \rangle = \vert \psi_1 \rangle \otimes \vert
\psi_{23} \rangle $ where 
the second and third qubits are in an entangled state $
\vert \psi_{23} \rangle $. The other two possible 
bipartitions are $
2 \vert 13 $ and $ 3 \vert 12 $. 
In case $ \vert \Psi \rangle  $ cannot be
written as either a fully separable or 
a biseparable state, then the state is said to 
possess genuine tripartite entanglement. 
There are two SLOCC inequivalent classes of genuine 
three-qubit
entanglement~\citep{dur-pra-00} namely, 
the GHZ and the W class. Hence any three-qubit
pure state can belong to either of the six SLOCC inequivalent 
classes \ie GHZ,
W, three different bi-separable classes or 
the separable class of states~\citep{dur-pra-00}.

We briefly outline below the procedure detailed in reference
\cite{zhao-pra-13}, for three-qubit pure state 
entanglement classification.
The entanglement measure
concurrence~\citep{wootters-prl-98,wootters-qic-01}
was used to identify biseparable states.
The most general three-qubit pure state can be written as 
\begin{equation}
 \vert \Psi
\rangle= \sum _{i,j,k=0}^{1} a_{ijk}\vert ijk \rangle \quad
{\rm with}\quad  \sum
_{i,j,k=0}^{1} \vert a_{ijk} \vert ^2 =1.
\end{equation}
The concurrence for state $ \rho=\vert
\Psi \rangle \langle \Psi \vert $ in the $1\vert 2 3$
partition is given 
by $ C(\rho)=\sqrt{1-(tr\rho_1)^2} $
where $ \rho_1=tr_2(\rho) $ is the reduced density 
operator of the first party.
The squared concurrence for a three-qubit pure state 
under the bipartition $ 1 \vert 23$ is given by
\begin{equation}
\label{conr}
C^2_{1 \vert 23}(\rho)=\sum \limits _{j,k=0}^{1} \vert a_{0jk} \vert^2
. \sum \limits _{j,k=0}^{1} \vert a_{1jk} \vert^2  
 -\Big\vert \sum \limits _{j,k=0}^{1} a_{0jk}a_{1jk}^* \Big\vert^2
\end{equation}

After a lengthy calculation, 
it was shown in \citep{zhao-pra-13} that 
the squared concurrence (Eq.~\ref{conr}) can be written as a quadratic
polynomial of the expectation values of Pauli operators for 
three qubits.
Using the symbol $ G_1(\rho) $ to denote
$ C^2_{1 \vert 23}(\rho) $, it  
takes the form
\begin{eqnarray}
\label{G1}
G_1(\rho)=&\frac{1}{16}&(3 - \langle\sigma_0\sigma_0\sigma_3 \rangle^2 
- \langle\sigma_0\sigma_3\sigma_0 \rangle^2 + \langle\sigma_3\sigma_3\sigma_0
  \rangle^2 \nonumber \\ -3 \langle\sigma_3\sigma_0 &\sigma_0 & \rangle^2  +
\langle\sigma_3\sigma_0\sigma_3 \rangle^2 - \langle\sigma_0\sigma_3\sigma_3
\rangle^2 + \langle\sigma_3\sigma_3\sigma_3 \rangle^2 \nonumber\\
- 3\langle\sigma_1\sigma_0 &\sigma_0 & \rangle^2 +
  \langle\sigma_1\sigma_0\sigma_3 \rangle^2 + \langle\sigma_1\sigma_3\sigma_0
\rangle^2 + \langle\sigma_1\sigma_3\sigma_3 \rangle^2 \nonumber\\
- 3\langle\sigma_2\sigma_0 &\sigma_0 & \rangle^2 +
  \langle\sigma_2\sigma_0\sigma_3 \rangle^2 + \langle\sigma_2\sigma_3\sigma_0
\rangle^2 + \langle\sigma_2\sigma_3\sigma_3 \rangle^2) \nonumber \\
\end{eqnarray}
with $ \sigma_0=\vert 0 \rangle\langle 0 \vert+\vert 1 \rangle\langle 1 \vert
$, $ \sigma_1=\vert 0 \rangle\langle 1 \vert+\vert 1 \rangle\langle 0 \vert  $,
$ \sigma_2=i(\vert 1 \rangle\langle 0 \vert-\vert 0 \rangle\langle 1 \vert)  $
and $ \sigma_3=\vert 0 \rangle\langle 0 \vert-\vert 1 \rangle\langle 1 \vert  $
being Pauli matrices in the computational basis. Similar expressions for
squared concurrences under the other two bipartitions \ie $ C^2_{2 \vert
13}(\rho) $ and $ C^2_{3 \vert 12}(\rho) $ can be written by permutation and
are symbolized by $ G_2(\rho) $ and $ G_3(\rho) $ respectively.

As described in \textit{Theorem 1} of \citep{zhao-pra-13}, 
for any three-qubit pure state $ \rho= \vert \Psi \rangle
\langle \Psi \vert $,
\begin{itemize}
\item[(i)]
$ \vert \Psi \rangle $ is fully separable iff $ G_l(\rho)=0
$, for $ l=2,3 $ or $ l=1,2 $ or $ l=1,3 $.
\item[(ii)]
 $ \vert \Psi \rangle $ is separable between \textit{l}$\rm
^{th}$ qubit and rest  iff $ G_l(\rho)=0 $ and $
G_m(\rho)>0$ with $ l,m\in \{ 1,2,3 \} $ and $ l\neq m $.
\item[(iii)]
$ \vert \Psi \rangle $ is genuinely entangled iff $
G_l(\rho)>0 $, for $ l=2,3 $ or $ l=1,2 $ or $ l=1,3 $.
\end{itemize}

Hence computing the entanglement witnesses $
G_l(\rho) $, through experimentally measured expectation
values of Pauli operators for an arbitrary three-qubit pure
state $ \rho= \vert \Psi \rangle \langle \Psi \vert $, can
immediately reveal the entanglement class of the state.

As per \textit{Theorem 1}-(iii) the current entanglement
classification protocol enables us to decide if a given pure
state has genuine three-qubit entanglement but does not
specify if the state belongs to the GHZ or the W class. To
overcome this limitation, we utilized our previous results
\citep{singh-arxiv-18} and define the observable
$O=2\sigma^{}_{1}\sigma^{}_{1}\sigma^{}_{1}$ and use the
$n$-tangle introduced in \cite{wong-pra-01,li-qip-12} as an
entanglement measure. For three qubits, a non-vanishing
3-tangle $\tau$, implies the state belongs to the GHZ class.
One may easily verify that for a given generic state $ \vert
\Psi \rangle $, the 3-tangle \ie $ \tau_{\Psi}=\langle \Psi
\vert O \vert \Psi \rangle ^2/4$.  Having defined $ O $ in
addition to $ G_l(\rho) $, the protocol is now equipped to
experimentally classify any three-qubit pure state. 
\subsection{Framework for Experimental Implementation}
It has been established~\citep{acin-prl-01} that any
three-qubit pure state can be transformed to a generic state
of the canonical form
\begin{equation}
\label{generic}
\vert\psi\rangle=a_0\vert 000 \rangle + a_1e^{\iota
\theta}\vert 100 \rangle + a_2\vert 101 \rangle + a_3\vert
110 \rangle + a_4\vert 111 \rangle
\end{equation}
where $a_i\geq 0$, $\sum_i a^2_i=1$ and $\theta \in
[0,\pi]$.  It should be noted that the entanglement
classification procedure outlined in Section~\ref{Theory}
works for any three-qubit pure state but we chose to
experimentally test it on arbitrary generic states, since
different states may have the same generic canonical
representation \citep{acin-prl-01}.  Entanglement properties
for the class of all such states can be fully characterized
resorting only to the SLOCC equivalent generic state
representative of that class. Such a choice of states
further eases the experimental implementation, as nearly
40\% of the expectation values of the Pauli operators
appearing in the expressions of $ G_l(\rho) $ (\textit{e.g.}
Eq.~\ref{G1}) vanish in the case of generic states
(Eq.~\ref{generic}).  
This entanglement classification protocol is not limited to generic states
but also works for any arbitrary three-qubit pure state of
form $ \vert \Psi \rangle= \sum _{i,j,k=0}^{1} a_{ijk}\vert
ijk \rangle $.
\subsection{Experimental Measurement of Observables using NMR}
\label{NMR-measurement}
We use nuclear magnetic resonance (NMR) hardware to
experimentally demonstrate the entanglement classification
protocol.  The crux of the detection protocol lies in
experimentally determining the expectation values of the
observables appearing in Eq.~\ref{G1}. In order to
experimentally find the expectation value of an observable
it is a standard practice to decompose it as a linear
superposition of some physically realizable basis operators.
One such widely used operator basis is the Pauli
basis~\citep{nielsen-book-02,oliveira-book-07}.  The next
step is to map the desired basis operator expectation value
to the experimentally accessible expectation value.  In NMR
the experimentally accessible information is the expectation
value  of Pauli $ z $-operator for each qubit.  We have
previously developed and demonstrated such a mapping
\citep{singh-pra-16,singh-arxiv-18} for any observable in
NMR. 

\begin{figure}[h]
\includegraphics[angle=0,scale=1]{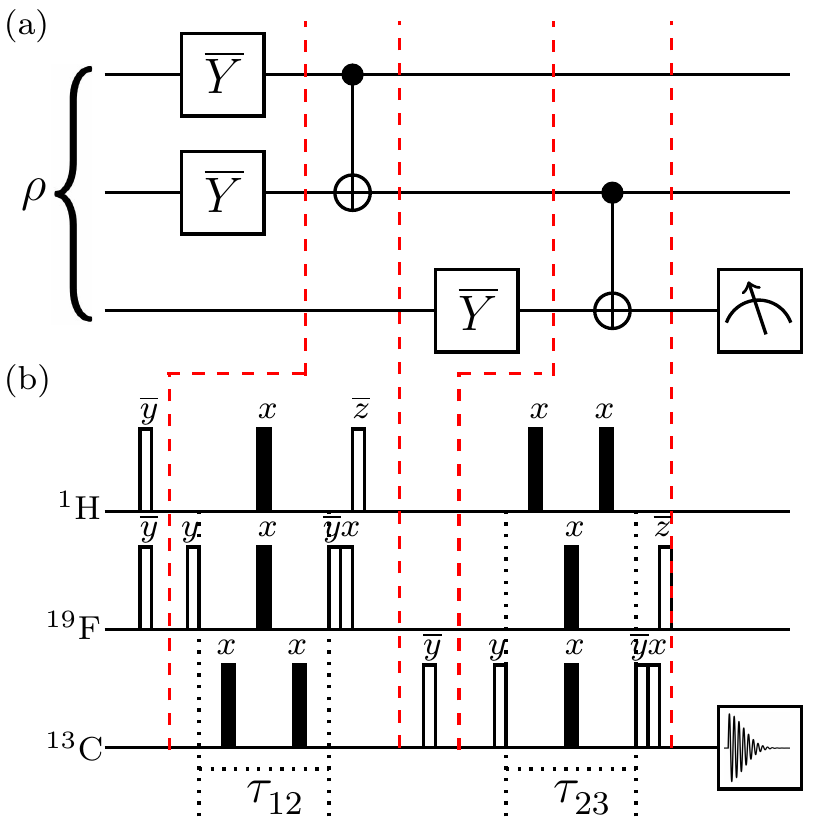}
\caption{(a) Quantum circuit to achieve mapping of the state
$ \rho $ to $ \rho_i $
followed by measurement of qubit 3 in the computational basis. (b) NMR 
pulse sequence to implement the quantum circuit given in (a). 
The unfilled rectangles denote $ \frac{\pi}{2} $
spin-selective RF pulses while the filled rectangles 
denote $\pi$ pulses. Pulse phases are written above the respective pulse 
and a bar over a phase represents negative phase. Delays are given 
by $\tau^{}_{ij}=1/(8 J_{ij})$; $i,j$ label the qubit and 
$J$ denotes the scalar coupling constant.}
\label{ckt+seq}
\end{figure}

Assuming that we are interested in the expectation value of
the operator $ O $ in the state $ \rho=\vert \Psi \rangle
\langle \Psi \vert $. To measure this we experimentally map
the state $ \rho \rightarrow \rho_i $ via map $
\rho_i=U_i^{\dagger}.\rho. U_i $ followed by measuring the
expectation value of Pauli $z$-operator in $ \rho_i $.
Explicit forms of $ U_i $ for two and three qubit systems
are given in \citep{singh-pra-16} and
\citep{singh-arxiv-18}, respectively. It can be easily
verified that $ \langle O \rangle $ in $ \rho $ is equal to
$ \langle \sigma_3 \rangle $ in $ \rho_i $ with $
U_i={\rm CNOT}_{23}.\overline{Y}_3.{\rm CNOT}_{12}.\overline{Y}_2.\overline{Y}_1
$. Here $ {\rm CNOT}_{ij} $ is the controlled-NOT gate with $ i $
as the control qubit and $ j $ as the target qubit. $ X(Y) $
are the local $ \frac{\pi}{2} $ unitary rotation having
phase $ x(y) $. Bar over a phase represents negative phase.
For the case of  $ \langle O \rangle $ in state $ \rho $ a
quantum circuit to achieve the state mapping is shown in
Fig.~\ref{ckt+seq} (a).

\begin{figure}[h]
\includegraphics[angle=0,scale=1.0]{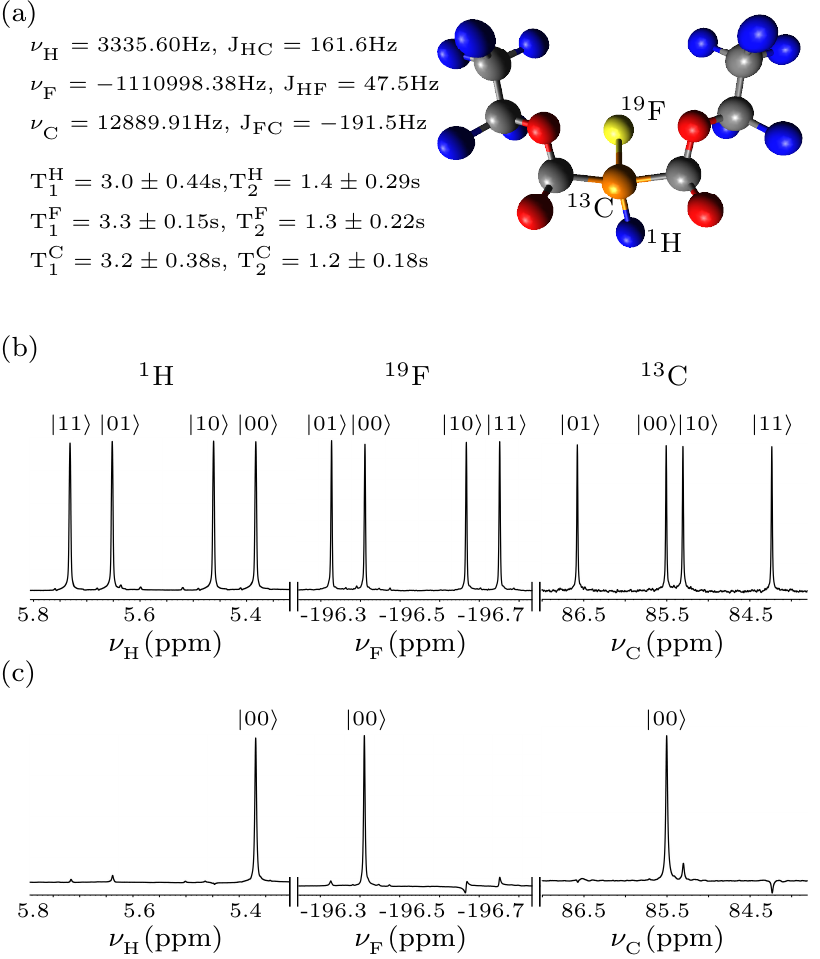}
\caption{(a) Molecular structure of
$^{13}$C-labeled diethyl fluoromalonate and NMR parameters.
NMR spectra of (b) thermal equilibrium state and (c) pseudopure
state. Each peak is labeled with the logical state of the 
passive qubit during the transition.}
\label{molecule} 
\end{figure}
\section{NMR Implementation of Three Qubit Entanglement Classification Protocol}
\label{NMR Implementation}
The Hamiltonian~\citep{ernst-book-90} in frequency units, for 
three qubits in the rotating frame can be written as
\begin{equation}\label{Hamiltonian}
\mathcal{H}= -\sum_{i=1}^{3} \nu_i I_{iz} + \sum_{i
>j,i=1}^{3} J_{ij}I_{iz}I_{jz}
\end{equation}
where the indices $i,j=$ 1,2 or 3 are the qubit labels,
$\nu_i$ is the chemical shift,
$J_{ij}$ is the scalar coupling constant, and $I_{iz} $ is the 
$z$-spin angular momentum operator of the
$i^{\rm{th}}$ spin.

For the experimental implementation of the entanglement classification
protocol, $^{13}$C labeled diethylfluoromalonate dissolved in acetone-D6 sample
was used, with the $^{1}$H, $^{19}$F and $^{13}$C nuclei serving as qubit 1,
qubit 2 and qubit 3, respectively.  Before preparing 
arbitrary three-qubit
pure states, the system was initialized in the pseudopure (PPS) state $\vert
000 \rangle$ utilizing spatial averaging~\cite{cory-physD-98}
with the PPS density operator given by
\begin{equation}
\rho_{000}=\frac{1-\epsilon}{2^3}\mathbb{I}_8 +\epsilon
\vert 000 \rangle \langle 000 \vert 
\end{equation} 
where $\epsilon \sim 10^{-5}$ is the thermal magnetic polarization at
room temperature and $ \mathbb{I}_8 $ is the 8 $ \times $ 8 identity operator.
The experimental
NMR parameters (rotating frame chemical shifts,
T$_1$ and T$_2$ relaxation times and 
scalar couplings $\rm{J}_{ij}$) 
as well as the NMR spectra of the thermal equilibrium and PPS states are shown in Fig.~\ref{molecule}. 
Each spectral transition in the NMR spectrum is labeled with the logical states of the passive qubits (\ie qubits not undergoing any transition) 
in the computational basis. Experimentally prepared PPS had fidelity (Fig.~\ref{molecule}(c)) 0.98$\pm$0.01 and was computed
using the fidelity measure
\citep{uhlmann-rpmp-76,jozsa-jmo-94}
\begin{equation}\label{fidelity}
F=\left[Tr\left(
\sqrt{\sqrt{\rho_{{\rm th}}}\rho_{{\rm ex}}
\sqrt{\rho_{{\rm th}}}}\right)\right]^2
\end{equation}
where $\rho_{{\rm th}}$ and $\rho_{{\rm ex}}$ are the theoretically
expected and the experimentally reconstructed density
operators, respectively. Fidelity measure is normalized in the sense that $ F\rightarrow 1 $ as
$\rho_{ex}\rightarrow\rho_{th}$. Experimental reconstruction of the density operator was achieved via full quantum state
tomography (QST)\citep{leskowitz-pra-04,singh-pla-16} utilizing 
a preparatory pulse set of $\left\lbrace III,
XXX, IIY, XYX, YII, XXY, IYY \right\rbrace$, where
$I$ implies ``no operation''. In NMR 
a $\frac{\pi}{2} $ local unitary rotation $X$($Y$) can be
achieved using highly accurate and calibrated  spin-selective transverse radio frequency
(RF) pulses having phase $x$($y$). 
\begin{figure}[h]
\includegraphics[angle=0,scale=1]{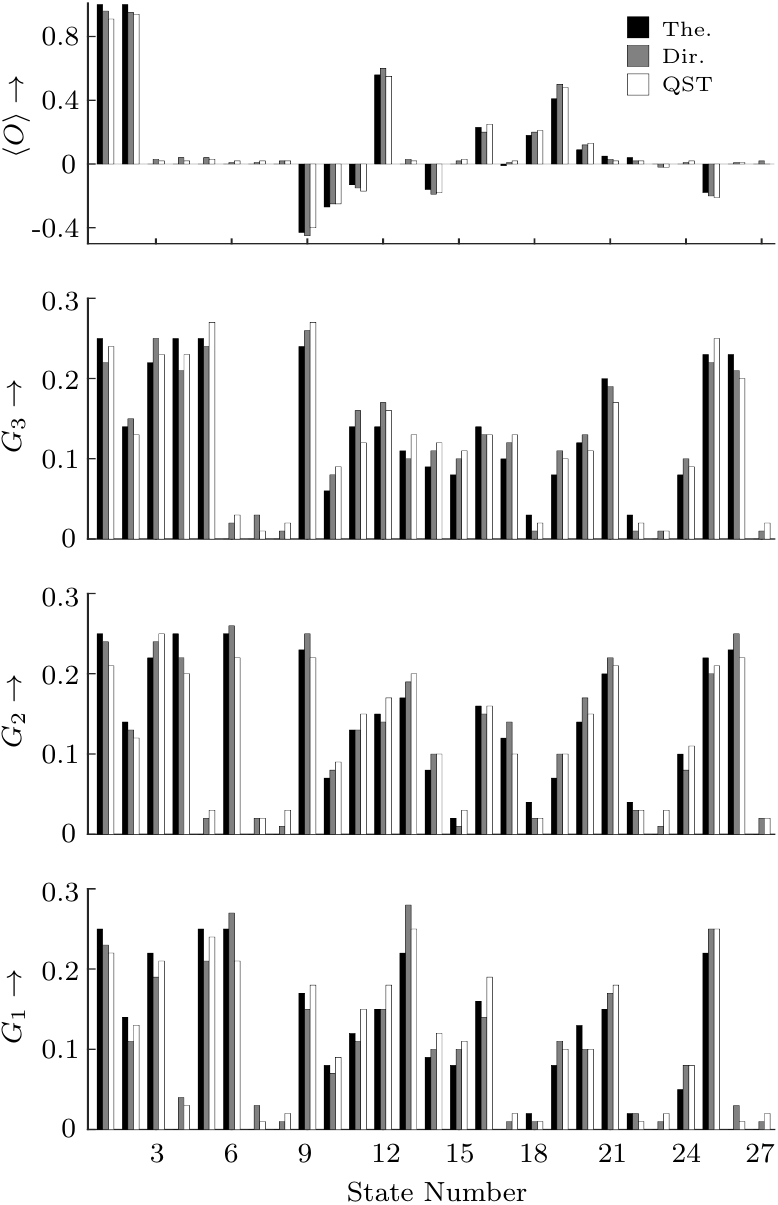}
\caption{Bar plots of the expectation 
value of the
observable $O$ and the squared 
concurrences $G_1$, $G_2$ and $G_3$ for states numbered
from 1-27 (Table~\ref{result table}). The 
state number is represented on the
horizontal axes
while the values of the respective observable
are represented along the vertical axes. Black, 
gray and
unfilled bars represent the theoretical (The.) values, directly
experimentally measured values (Dir.), and 
QST-derived values, respectively.} 
\label{ResultPlot}
\end{figure}

\begin{table*} [t]
\caption{\label{result table}
Results of entanglement classification protocol 
for twenty seven states. Label BS 
denotes a biseparable state while R 
denotes a randomly prepared state. 
The first column depicts the state label, 
the top row lists the observable (Obs.) while 
the second row specifies if the 
observable value obtained is theoretical (The.), 
from QST or direct experimentally determined (Dir.).}
\begin{ruledtabular}
\begin{tabular}{c | c c c | c c c | c c c | c c c}

Obs. $\rightarrow$ &  \multicolumn{3}{c}{$ \langle O \rangle $} & \multicolumn{3}{c}{$G_1$} & \multicolumn{3}{c}{$G_2$} &
\multicolumn{3}{c}{$G_3$} \\

State (F) $\downarrow$ & The. & QST & Dir. & The. & QST & Dir. & The. & QST & Dir. & The. & QST & Dir.  \\
\colrule
GHZ(0.96$ \pm $0.01) & 1.00 & 0.96 & 0.91 & 0.25 & 0.23 & 0.22 & 0.25 & 0.24 & 0.21 & 0.25 & 0.22 & 0.24 \\
$\rm W\overline{W}$(0.95$ \pm $0.02)  & 1.00 & 0.95 & 0.94 & 0.14 & 0.11 & 0.13 & 0.14 & 0.13 & 0.12 & 0.14 & 0.15 & 0.13 \\
W(0.96$ \pm $0.02) & 0 & 0.03 & 0.02 & 0.22 & 0.19 & 0.21 & 0.22 & 0.24 & 0.25 & 0.22 & 0.25 & 0.23 \\
BS$_1$(0.98$ \pm $0.01) & 0 & 0.04 & 0.02 & 0 & 0.04 & 0.03 & 0.25 & 0.22 & 0.20 & 0.25 & 0.21 & 0.23 \\
BS$_2$(0.94$ \pm $0.03) & 0 & 0.04 & 0.03 & 0.25 & 0.21 & 0.24 & 0 & 0.02 & 0.03 & 0.25 & 0.24 & 0.27 \\
BS$_3$(0.95$ \pm $0.02) & 0 & 0.01 & 0.02 & 0.25 & 0.27 & 0.21 & 0.25 & 0.26 & 0.22 & 0 & 0.02 & 0.03 \\
Sep(0.98$ \pm $0.01) & 0 & 0.01 & 0.02 & 0 & 0.03 & 0.01 & 0 & 0.02 & 0.02 & 0 & 0.03 & 0.01 \\
R$_{ 1}$(0.92$ \pm $0.03) & 0 & 0.02 & 0.02 & 0 & 0.01 & 0.02 & 0 & 0.01 & 0.03 & 0 & 0.01 & 0.02 \\
R$_{ 2}$(0.93$ \pm $0.02) & -0.43 & -0.45 & -0.40 & 0.17 & 0.15 & 0.18 & 0.23 & 0.25 & 0.22 & 0.24 & 0.26 & 0.27 \\
R$_{ 3}$(0.96$ \pm $0.02) & -0.27 & -0.25 & -0.25 & 0.08 & 0.07 & 0.09 & 0.07 & 0.08 & 0.09 & 0.06 & 0.08 & 0.09 \\
R$_{ 4}$(0.94$ \pm $0.03) & -0.13 & -0.15 & -0.17 & 0.12 & 0.11 & 0.15 & 0.13 & 0.13 & 0.15 & 0.14 & 0.16 & 0.12 \\
R$_{ 5}$(0.93$ \pm $0.02) & 0.56 & 0.60 & 0.55 & 0.15 & 0.15 & 0.18 & 0.15 & 0.14 & 0.17 & 0.14 & 0.17 & 0.16 \\
R$_{ 6}$(0.89$ \pm $0.01) & 0 & 0.03 & 0.02 & 0.22 & 0.28 & 0.25 & 0.17 & 0.19 & 0.20 & 0.11 & 0.10 & 0.13 \\
R$_{ 7}$(0.96$ \pm $0.02) & -0.16 & -0.19 & -0.18 & 0.09 & 0.10 & 0.12 & 0.08 & 0.10 & 0.10 & 0.09 & 0.11 & 0.12 \\
R$_{ 8}$(0.93$ \pm $0.02) & 0 & 0.02 & 0.03 & 0.08 & 0.10 & 0.11 & 0.02 & 0.01 & 0.03 & 0.08 & 0.10 & 0.11 \\
R$_{ 9}$(0.97$ \pm $0.03) & 0.23 & 0.20 & 0.25 & 0.16 & 0.14 & 0.19 & 0.16 & 0.15 & 0.16 & 0.14 & 0.13 & 0.13 \\
R$_{10}$(0.93$ \pm $0.02) & -0.01 & 0.01 & 0.02 & 0 & 0.01 & 0.02 & 0.12 & 0.14 & 0.10 & 0.10 & 0.12 & 0.13 \\
R$_{11}$(0.94$ \pm $0.01) & 0.18 & 0.20 & 0.21 & 0.02 & 0.01 & 0.01 & 0.04 & 0.02 & 0.02 & 0.03 & 0.01 & 0.02 \\
R$_{12}$(0.95$ \pm $0.02) & 0.41 & 0.50 & 0.48 & 0.08 & 0.11 & 0.10 & 0.07 & 0.10 & 0.10 & 0.08 & 0.11 & 0.10 \\
R$_{13}$(0.93$ \pm $0.01) & 0.09 & 0.12 & 0.13 & 0.13 & 0.10 & 0.10 & 0.14 & 0.17 & 0.15 & 0.12 & 0.13 & 0.11 \\
R$_{14}$(0.94$ \pm $0.02) & 0.05 & 0.03 & 0.02 & 0.15 & 0.17 & 0.18 & 0.20 & 0.22 & 0.21 & 0.20 &  0.19 & 0.17 \\
R$_{15}$(0.98$ \pm $0.01) & 0.04 & 0.02 & 0.02 & 0.02 & 0.02 & 0.01 & 0.04 & 0.03 & 0.03 & 0.03 & 0.01 & 0.02 \\
R$_{16}$(0.96$ \pm $0.01) & 0 & -0.02 & -0.02 & 0 & 0.01 & 0.02 & 0 & 0.01 & 0.03 & 0 & 0.01 & 0.01 \\
R$_{17}$(0.95$ \pm $0.02) & 0 & 0.01 & 0.02 & 0.05 & 0.08 & 0.08 & 0.10 & 0.08 & 0.11 & 0.08 & 0.10 & 0.09 \\
R$_{18}$(0.90$ \pm $0.02) & -0.18 & -0.20 & -0.21 & 0.22 & 0.25 & 0.25 & 0.22 & 0.20 & 0.21 & 0.23 & 0.22 & 0.25 \\
R$_{19}$(0.94$ \pm $0.02) & 0 & 0.01 & 0.01 & 0 & 0.03 & 0.01 & 0.23 & 0.25 & 0.22 & 0.23 & 0.21 & 0.20 \\
R$_{20}$(0.96$ \pm $0.02) & 0 & 0.02 & 0 & 0 & 0.01 & 0.02 & 0 & 0.02 & 0.02 & 0 & 0.01 & 0.02 \\

\end{tabular}
\end{ruledtabular}
\end{table*}

A Bruker Avance-III 600-MHz FT-NMR spectrometer equipped with a QXI probe was
used for the experiments which were performed at 
room temperature
($293$K). A spin specific pulse calibration yields the duration, amplitude and
phase to achieve the desired local unitary operation. Free evolution under the
Hamiltonian Eq.~\ref{Hamiltonian} for a desired duration was used to achieve
non-local unitary operations. $\frac{\pi}{2}$ spin selective pulses for
$^{1}$H, $^{19}$F and $^{13}$C in the current study were 9.40 $\mu$s at 18.14 W
power level, 22.50 $\mu$s at a power level of 42.27 W and 16.00 $\mu$s at a
power level of  179.47 W, respectively.

For the experimental demonstration of the entanglement classification protocol
discussed in Section~\ref{Theory}, we prepared the three qubits in twenty
seven different states.  Seven states were prepared from the six SLOCC
inequivalent entanglement classes \ie  GHZ (GHZ and $\rm W\overline{W}$
states), W, three bi-separable and from the separable class of states. 
We labeled
three biseparable class states under partitions $ 1\vert 23 $, $ 2\vert 13 $
and $ 3\vert 12 $ as BS$ _1 $, BS$ _2 $ and BS$ _3 $ respectively.
Additionally, twenty random generic states were prepared 
using a random number generator and labeled as R$_1$,
R$_2$, R$_3$,......., R$_{20}$. 
The
details of the quantum circuits as
well as NMR pulse sequences required to prepare all the desired quantum states
in the current study are given in
Reference~\cite{dogra-pra-15}. All the prepared states were found to
have the fidelity (F)
in the range 0.88 to 0.99. For each prepared state the expectation values
of the Pauli operators were determined as 
described in Section~\ref{NMR-measurement}
which in turn was used to compute $ G_l(\rho) $ using Eq.~\ref{G1}. $ \langle O
\rangle $ was also found in all the cases as it serves as an entanglement
witness of the GHZ class.

Experimental results of the three-qubit entanglement classification and
detection protocol are shown in Table~\ref{result table}.  A bar chart has been
plotted in Fig.~\ref{ResultPlot} for a visual representation of the
experimental results of Table~\ref{result table}. To obtain the bar plots of
Fig.~\ref{ResultPlot}, the experimentally prepared states were numbered from 1
to 27 as per the ordering in Table~\ref{result table}. As detailed in
Sec.~\ref{Theory}, the concurrence $ G_l(\rho) $ acts as an entanglement
witness, and the additional observable $ O $ helps in the experimental
discrimination of GHZ class states from the rest. In order to validate the
experimental results we also computed the negativity
\citep{weinstein-pra-10,vidal-pra-02} from the experimentally reconstructed
state via QST \citep{leskowitz-pra-04} and the results are shown in
Table~\ref{negativity table}. In each case, the experiments were repeated
several times for experimental error estimates. Experimental errors were in the
range of 2.2\% - 5.7\% for the values reported in the Table~\ref{result table}.  

As observed from Table~\ref{result table}, the seven states, from six SLOCC
inequivalent classes, were prepared with experimental fidelity $ \geq $ 0.95.
The entanglement classes of all these seven states were correctly identified
with the current protocol. Further, the states R$ _2 $, R$ _3 $, R$ _4 $, R$ _5
$, R$ _6 $, R$ _7 $, R$ _8 $, R$ _9 $, R$ _{11} $, R$ _{12} $, R$ _{13} $, R$
_{14} $, R$ _{17} $ and R$ _{18} $ have at least two non-zero concurrences and
hence are genuinely entangled states.  This fact is further supported by
negativity of these states as reported in Table~\ref{negativity table}. As
discussed earlier, in order to discriminate GHZ class from the rest one can
resort to the observable $ O $. Non vanishing values of $ \langle O \rangle $
in Table~\ref{result table} imply that the states R$ _2 $, R$ _3 $, R$ _4 $, R$
_5 $, R$ _7 $, R$ _9 $, R$ _{11} $, R$ _{12} $, R$ _{13} $ and R$ _{18} $
belong to the GHZ class. In contrast, the genuinely entangled states R$ _6 $,
R$ _8 $, R$ _{14} $ and R$ _{17} $ have vanishing values of $ \langle O \rangle
$ and hence have vanishing 3-tangle as well, so they were identified as
belonging to the W class.  States R$ _{10} $ and R$ _{19} $ have vanishing
concurrence $ G_1 $ implying that these states belong to BS$ _1 $ class.  Also
states R$ _{1} $, R$ _{15} $, R$ _{16} $ and R$ _{20} $ were identified as
separable as all the observables have near zero values as well as zero
negativity.

\begin{table}[t]
\caption{\label{negativity table}
Theoretically calculated and experimentally 
measured negativity values for all 
twenty seven states under investigation.
}
\begin{ruledtabular}
\begin{tabular}{c | c  c}
Negativity $\rightarrow$ &  Theoretical  & Experimental \\
State $\downarrow$ &  &   \\
\colrule
GHZ & 0.5 & 0.47 $\pm$ 0.02 \\
$\rm W\overline{W}$ & 0.37 & 0.39 $\pm$ 0.02 \\
W       & 0.47 & 0.44 $\pm$ 0.01\\
BS$_1$  & 0    & 0.02 $\pm$ 0.02\\
BS$_2$  & 0    & 0.03 $\pm$ 0.01\\
BS$_3$  & 0    & 0.02 $\pm$ 0.02\\
Sep     & 0    & 0.02 $\pm$ 0.02\\
R$_{1}$ & 0    & 0.01 $\pm$ 0.01\\
R$_{2}$ & 0.46 & 0.43 $\pm$ 0.04\\
R$_{3}$ & 0.26 & 0.24 $\pm$ 0.03\\
R$_{4}$ & 0.18 & 0.17 $\pm$ 0.03\\
R$_{5}$ & 0.38 & 0.35 $\pm$ 0.02\\
R$_{6}$ & 0.40 & 0.37 $\pm$ 0.04\\
R$_{7}$ & 0.29 & 0.31 $\pm$ 0.03\\
R$_{8}$ & 0.22 & 0.21 $\pm$ 0.02\\
R$_{9}$ & 0.39 & 0.37 $\pm$ 0.04\\
R$_{10}$ &0.03 & 0.01 $\pm$ 0.01\\
R$_{11}$ &0.17 & 0.14 $\pm$ 0.02\\
R$_{12}$ &0.27 & 0.30 $\pm$ 0.03\\
R$_{13}$ &0.16 & 0.12 $\pm$ 0.04\\
R$_{14}$ &0.42 & 0.37 $\pm$ 0.04\\
R$_{15}$ &0.02 & 0.03 $\pm$ 0.01\\
R$_{16}$ &0    & 0.01 $\pm$ 0.01\\
R$_{17}$ &0.26 & 0.22 $\pm$ 0.03\\
R$_{18}$ &0.47 & 0.41 $\pm$ 0.04\\
R$_{19}$ &0    & 0.02 $\pm$ 0.02\\
R$_{20}$ &0    & 0.03 $\pm$ 0.02\\
\end{tabular}
\end{ruledtabular}
\end{table}
\section{Concluding Remarks} 
\label{remarks} 
We experimentally classified the entanglement of
arbitrary three-qubit pure states using the
concurrence as an entanglement measure.  Concurrence was measured
experimentally by measuring the expectation values of the
Pauli operators under all three bipartitions. To
demonstrate the efficacy of the
entanglement classification scheme experimentally, we
tested the protocol on 
seven
standard as well as twenty random three-qubit 
pure states prepared on an NMR quantum information
processor.
The entanglement class of all the seven states representing six
SLOCC classes was correctly identified.  The results were
validated using full QST and negativity calculations for
each state. The entanglement class of the twenty random states
was also identified within experimental error limits.
Non-zero negativity as well as two out of three concurrence
witnesses indicated that the state under investigation had
genuine three-qubit entanglement. Such states may belong to
either the GHZ or the W class. To further differentiate the entanglement
class, we measured the three-tangle in each case, since non-zero
three-tangle is a signature of the GHZ class. Based on this,
the experimental classification protocol successfully
classified randomly generated states  
as belonging to either the GHZ class or W class of entangled
states or as biseparable or separable states.
Future directions of this
work include evaluating the performance of the protocol for
mixed states as well as for larger qubit registers.
\begin{acknowledgments}
All the experiments were performed on a Bruker Avance-III
600 MHz FT-NMR spectrometer at the NMR Research Facility of
IISER Mohali. Arvind acknowledges funding from DST India
under Grant No. EMR/2014/000297. K.D. acknowledges funding
from DST India under Grant No. EMR/2015/000556.
\end{acknowledgments}

\begin{thebibliography}{56}%
\makeatletter
\providecommand \@ifxundefined [1]{%
 \@ifx{#1\undefined}
}%
\providecommand \@ifnum [1]{%
 \ifnum #1\expandafter \@firstoftwo
 \else \expandafter \@secondoftwo
 \fi
}%
\providecommand \@ifx [1]{%
 \ifx #1\expandafter \@firstoftwo
 \else \expandafter \@secondoftwo
 \fi
}%
\providecommand \natexlab [1]{#1}%
\providecommand \enquote  [1]{``#1''}%
\providecommand \bibnamefont  [1]{#1}%
\providecommand \bibfnamefont [1]{#1}%
\providecommand \citenamefont [1]{#1}%
\providecommand \href@noop [0]{\@secondoftwo}%
\providecommand \href [0]{\begingroup \@sanitize@url \@href}%
\providecommand \@href[1]{\@@startlink{#1}\@@href}%
\providecommand \@@href[1]{\endgroup#1\@@endlink}%
\providecommand \@sanitize@url [0]{\catcode `\\12\catcode `\$12\catcode
  `\&12\catcode `\#12\catcode `\^12\catcode `\_12\catcode `\%12\relax}%
\providecommand \@@startlink[1]{}%
\providecommand \@@endlink[0]{}%
\providecommand \url  [0]{\begingroup\@sanitize@url \@url }%
\providecommand \@url [1]{\endgroup\@href {#1}{\urlprefix }}%
\providecommand \urlprefix  [0]{URL }%
\providecommand \Eprint [0]{\href }%
\@ifxundefined \urlstyle {%
  \providecommand \doi  [0]{\begingroup \@sanitize@url \@doi}%
  \providecommand \@doi [1]{\endgroup \@@startlink {\doibase
  #1}doi:\discretionary {}{}{}#1\@@endlink }%
}{%
  \providecommand \doi  [0]{doi:\discretionary{}{}{}\begingroup
  \urlstyle{rm}\Url }%
}%
\providecommand \doibase [0]{http://dx.doi.org/}%
\providecommand \Doi [0]{\begingroup \@sanitize@url \@Doi }%
\providecommand \@Doi  [1]{\endgroup\@@startlink{\doibase#1}\@@Doi}%
\providecommand \@@Doi [1]{#1\@@endlink}%
\providecommand \selectlanguage [0]{\@gobble}%
\providecommand \bibinfo  [0]{\@secondoftwo}%
\providecommand \bibfield  [0]{\@secondoftwo}%
\providecommand \translation [1]{[#1]}%
\providecommand \BibitemOpen [0]{}%
\providecommand \bibitemStop [0]{}%
\providecommand \bibitemNoStop [0]{.\EOS\space}%
\providecommand \EOS [0]{\spacefactor3000\relax}%
\providecommand \BibitemShut  [1]{\csname bibitem#1\endcsname}%
\bibitem [{\citenamefont {Horodecki}\ \emph {et~al.}(2009)\citenamefont
  {Horodecki}, \citenamefont {Horodecki}, \citenamefont {Horodecki},\ and\
  \citenamefont {Horodecki}}]{horodecki-rmp-09}%
  \BibitemOpen
  \bibfield  {author} {\bibinfo {author} {\bibfnamefont {R.}~\bibnamefont
  {Horodecki}}, \bibinfo {author} {\bibfnamefont {P.}~\bibnamefont
  {Horodecki}}, \bibinfo {author} {\bibfnamefont {M.}~\bibnamefont
  {Horodecki}}, \ and\ \bibinfo {author} {\bibfnamefont {K.}~\bibnamefont
  {Horodecki}},\ }\Doi {10.1103/RevModPhys.81.865} {\bibfield  {journal}
  {\bibinfo  {journal} {Rev. Mod. Phys.},\ }\textbf {\bibinfo {volume} {81}},\
  \bibinfo {pages} {865} (\bibinfo {year} {2009})}\BibitemShut {NoStop}%
\bibitem [{\citenamefont {G\"uhne}\ and\ \citenamefont
  {T\'oth}(2009)}]{guhne-pr-09}%
  \BibitemOpen
  \bibfield  {author} {\bibinfo {author} {\bibfnamefont {O.}~\bibnamefont
  {G\"uhne}}\ and\ \bibinfo {author} {\bibfnamefont {G.}~\bibnamefont
  {T\'oth}},\ }\Doi {10.1016/j.physrep.2009.02.004} {\bibfield  {journal}
  {\bibinfo  {journal} {Phys. Rep.},\ }\textbf {\bibinfo {volume} {474}},\
  \bibinfo {pages} {1 } (\bibinfo {year} {2009})}\BibitemShut {NoStop}%
\bibitem [{\citenamefont {Li}\ \emph {et~al.}(2013)\citenamefont {Li},
  \citenamefont {Zhao}, \citenamefont {Fei},\ and\ \citenamefont
  {Wang}}]{li-ff-13}%
  \BibitemOpen
  \bibfield  {author} {\bibinfo {author} {\bibfnamefont {M.}~\bibnamefont
  {Li}}, \bibinfo {author} {\bibfnamefont {M.-J.}\ \bibnamefont {Zhao}},
  \bibinfo {author} {\bibfnamefont {S.-M.}\ \bibnamefont {Fei}}, \ and\
  \bibinfo {author} {\bibfnamefont {Z.-X.}\ \bibnamefont {Wang}},\ }\Doi
  {10.1007/s11467-013-0355-3} {\bibfield  {journal} {\bibinfo  {journal}
  {Front. Phys.},\ }\textbf {\bibinfo {volume} {8}},\ \bibinfo {pages} {357}
  (\bibinfo {year} {2013})}\BibitemShut {NoStop}%
\bibitem [{\citenamefont {Thew}\ \emph {et~al.}(2002)\citenamefont {Thew},
  \citenamefont {Nemoto}, \citenamefont {White},\ and\ \citenamefont
  {Munro}}]{thew-pra-02}%
  \BibitemOpen
  \bibfield  {author} {\bibinfo {author} {\bibfnamefont {R.~T.}\ \bibnamefont
  {Thew}}, \bibinfo {author} {\bibfnamefont {K.}~\bibnamefont {Nemoto}},
  \bibinfo {author} {\bibfnamefont {A.~G.}\ \bibnamefont {White}}, \ and\
  \bibinfo {author} {\bibfnamefont {W.~J.}\ \bibnamefont {Munro}},\ }\Doi
  {10.1103/PhysRevA.66.012303} {\bibfield  {journal} {\bibinfo  {journal}
  {Phys. Rev. A},\ }\textbf {\bibinfo {volume} {66}},\ \bibinfo {pages}
  {012303} (\bibinfo {year} {2002})}\BibitemShut {NoStop}%
\bibitem [{\citenamefont {G\"uhne}\ \emph {et~al.}(2003)\citenamefont
  {G\"uhne}, \citenamefont {Hyllus}, \citenamefont {Bru\ss{}}, \citenamefont
  {Ekert}, \citenamefont {Lewenstein}, \citenamefont {Macchiavello},\ and\
  \citenamefont {Sanpera}}]{guhne-jmo-03}%
  \BibitemOpen
  \bibfield  {author} {\bibinfo {author} {\bibfnamefont {O.}~\bibnamefont
  {G\"uhne}}, \bibinfo {author} {\bibfnamefont {P.}~\bibnamefont {Hyllus}},
  \bibinfo {author} {\bibfnamefont {D.}~\bibnamefont {Bru\ss{}}}, \bibinfo
  {author} {\bibfnamefont {A.}~\bibnamefont {Ekert}}, \bibinfo {author}
  {\bibfnamefont {M.}~\bibnamefont {Lewenstein}}, \bibinfo {author}
  {\bibfnamefont {C.}~\bibnamefont {Macchiavello}}, \ and\ \bibinfo {author}
  {\bibfnamefont {A.}~\bibnamefont {Sanpera}},\ }\Doi
  {10.1080/09500340308234554} {\bibfield  {journal} {\bibinfo  {journal} {J.
  Mod. Optics},\ }\textbf {\bibinfo {volume} {50}},\ \bibinfo {pages} {1079}
  (\bibinfo {year} {2003})}\BibitemShut {NoStop}%
\bibitem [{\citenamefont {Arrazola}\ \emph {et~al.}(2012)\citenamefont
  {Arrazola}, \citenamefont {Gittsovich},\ and\ \citenamefont
  {L\"utkenhaus}}]{arrazola-pra-12}%
  \BibitemOpen
  \bibfield  {author} {\bibinfo {author} {\bibfnamefont {J.~M.}\ \bibnamefont
  {Arrazola}}, \bibinfo {author} {\bibfnamefont {O.}~\bibnamefont
  {Gittsovich}}, \ and\ \bibinfo {author} {\bibfnamefont {N.}~\bibnamefont
  {L\"utkenhaus}},\ }\Doi {10.1103/PhysRevA.85.062327} {\bibfield  {journal}
  {\bibinfo  {journal} {Phys. Rev. A},\ }\textbf {\bibinfo {volume} {85}},\
  \bibinfo {pages} {062327} (\bibinfo {year} {2012})}\BibitemShut {NoStop}%
\bibitem [{\citenamefont {Jungnitsch}\ \emph {et~al.}(2011)\citenamefont
  {Jungnitsch}, \citenamefont {Moroder},\ and\ \citenamefont
  {G\"uhne}}]{jungnitsch-prl-11}%
  \BibitemOpen
  \bibfield  {author} {\bibinfo {author} {\bibfnamefont {B.}~\bibnamefont
  {Jungnitsch}}, \bibinfo {author} {\bibfnamefont {T.}~\bibnamefont {Moroder}},
  \ and\ \bibinfo {author} {\bibfnamefont {O.}~\bibnamefont {G\"uhne}},\ }\Doi
  {10.1103/PhysRevLett.106.190502} {\bibfield  {journal} {\bibinfo  {journal}
  {Phys. Rev. Lett.},\ }\textbf {\bibinfo {volume} {106}},\ \bibinfo {pages}
  {190502} (\bibinfo {year} {2011})}\BibitemShut {NoStop}%
\bibitem [{\citenamefont {Peres}(1996)}]{peres-prl-96}%
  \BibitemOpen
  \bibfield  {author} {\bibinfo {author} {\bibfnamefont {A.}~\bibnamefont
  {Peres}},\ }\Doi {10.1103/PhysRevLett.77.1413} {\bibfield  {journal}
  {\bibinfo  {journal} {Phys. Rev. Lett.},\ }\textbf {\bibinfo {volume} {77}},\
  \bibinfo {pages} {1413} (\bibinfo {year} {1996})}\BibitemShut {NoStop}%
\bibitem [{\citenamefont {Li}\ \emph {et~al.}(2017)\citenamefont {Li},
  \citenamefont {Wang}, \citenamefont {Shen}, \citenamefont {Chen},\ and\
  \citenamefont {Fei}}]{li-sr-17}%
  \BibitemOpen
  \bibfield  {author} {\bibinfo {author} {\bibfnamefont {M.}~\bibnamefont
  {Li}}, \bibinfo {author} {\bibfnamefont {J.}~\bibnamefont {Wang}}, \bibinfo
  {author} {\bibfnamefont {S.}~\bibnamefont {Shen}}, \bibinfo {author}
  {\bibfnamefont {Z.}~\bibnamefont {Chen}}, \ and\ \bibinfo {author}
  {\bibfnamefont {S.-M.}\ \bibnamefont {Fei}},\ }\Doi
  {10.1038/s41598-017-17585-7} {\bibfield  {journal} {\bibinfo  {journal} {Sc.
  Rep.},\ }\textbf {\bibinfo {volume} {7}},\ \bibinfo {pages} {17274} (\bibinfo
  {year} {2017})}\BibitemShut {NoStop}%
\bibitem [{\citenamefont {DiVincenzo}\ and\ \citenamefont
  {Peres}(1997)}]{peres-pra-97}%
  \BibitemOpen
  \bibfield  {author} {\bibinfo {author} {\bibfnamefont {D.~P.}\ \bibnamefont
  {DiVincenzo}}\ and\ \bibinfo {author} {\bibfnamefont {A.}~\bibnamefont
  {Peres}},\ }\Doi {10.1103/PhysRevA.55.4089} {\bibfield  {journal} {\bibinfo
  {journal} {Phys. Rev. A},\ }\textbf {\bibinfo {volume} {55}},\ \bibinfo
  {pages} {4089} (\bibinfo {year} {1997})}\BibitemShut {NoStop}%
\bibitem [{\citenamefont {Neeley}\ \emph {et~al.}(2010)\citenamefont {Neeley},
  \citenamefont {Bialczak}, \citenamefont {Lenander}, \citenamefont {Lucero},
  \citenamefont {Mariantoni}, \citenamefont {O'Connell}, \citenamefont {Sank},
  \citenamefont {Wang}, \citenamefont {Weides}, \citenamefont {Wenner},
  \citenamefont {Yin}, \citenamefont {Yamamoto}, \citenamefont {Cleland},\ and\
  \citenamefont {Martinis}}]{neeley-nature-10}%
  \BibitemOpen
  \bibfield  {author} {\bibinfo {author} {\bibfnamefont {M.}~\bibnamefont
  {Neeley}}, \bibinfo {author} {\bibfnamefont {R.~C.}\ \bibnamefont
  {Bialczak}}, \bibinfo {author} {\bibfnamefont {M.}~\bibnamefont {Lenander}},
  \bibinfo {author} {\bibfnamefont {E.}~\bibnamefont {Lucero}}, \bibinfo
  {author} {\bibfnamefont {M.}~\bibnamefont {Mariantoni}}, \bibinfo {author}
  {\bibfnamefont {A.~D.}\ \bibnamefont {O'Connell}}, \bibinfo {author}
  {\bibfnamefont {D.}~\bibnamefont {Sank}}, \bibinfo {author} {\bibfnamefont
  {H.}~\bibnamefont {Wang}}, \bibinfo {author} {\bibfnamefont {M.}~\bibnamefont
  {Weides}}, \bibinfo {author} {\bibfnamefont {J.}~\bibnamefont {Wenner}},
  \bibinfo {author} {\bibfnamefont {Y.}~\bibnamefont {Yin}}, \bibinfo {author}
  {\bibfnamefont {T.}~\bibnamefont {Yamamoto}}, \bibinfo {author}
  {\bibfnamefont {A.~N.}\ \bibnamefont {Cleland}}, \ and\ \bibinfo {author}
  {\bibfnamefont {J.~M.}\ \bibnamefont {Martinis}},\ }\href
  {http://dx.doi.org/10.1038/nature09418} {\bibfield  {journal} {\bibinfo
  {journal} {Nature},\ }\textbf {\bibinfo {volume} {467}},\ \bibinfo {pages}
  {570} (\bibinfo {year} {2010})}\BibitemShut {NoStop}%
\bibitem [{\citenamefont {Neumann}\ \emph {et~al.}(2008)\citenamefont
  {Neumann}, \citenamefont {Mizuochi}, \citenamefont {Rempp}, \citenamefont
  {Hemmer}, \citenamefont {Watanabe}, \citenamefont {Yamasaki}, \citenamefont
  {Jacques}, \citenamefont {Gaebel}, \citenamefont {Jelezko},\ and\
  \citenamefont {Wrachtrup}}]{neumann-science-08}%
  \BibitemOpen
  \bibfield  {author} {\bibinfo {author} {\bibfnamefont {P.}~\bibnamefont
  {Neumann}}, \bibinfo {author} {\bibfnamefont {N.}~\bibnamefont {Mizuochi}},
  \bibinfo {author} {\bibfnamefont {F.}~\bibnamefont {Rempp}}, \bibinfo
  {author} {\bibfnamefont {P.}~\bibnamefont {Hemmer}}, \bibinfo {author}
  {\bibfnamefont {H.}~\bibnamefont {Watanabe}}, \bibinfo {author}
  {\bibfnamefont {S.}~\bibnamefont {Yamasaki}}, \bibinfo {author}
  {\bibfnamefont {V.}~\bibnamefont {Jacques}}, \bibinfo {author} {\bibfnamefont
  {T.}~\bibnamefont {Gaebel}}, \bibinfo {author} {\bibfnamefont
  {F.}~\bibnamefont {Jelezko}}, \ and\ \bibinfo {author} {\bibfnamefont
  {J.}~\bibnamefont {Wrachtrup}},\ }\Doi {10.1126/science.1157233} {\bibfield
  {journal} {\bibinfo  {journal} {Science},\ }\textbf {\bibinfo {volume}
  {320}},\ \bibinfo {pages} {1326} (\bibinfo {year} {2008})}\BibitemShut
  {NoStop}%
\bibitem [{\citenamefont {Dogra}\ \emph {et~al.}(2015)\citenamefont {Dogra},
  \citenamefont {Dorai},\ and\ \citenamefont {Arvind}}]{dogra-pra-15}%
  \BibitemOpen
  \bibfield  {author} {\bibinfo {author} {\bibfnamefont {S.}~\bibnamefont
  {Dogra}}, \bibinfo {author} {\bibfnamefont {K.}~\bibnamefont {Dorai}}, \ and\
  \bibinfo {author} {\bibnamefont {Arvind}},\ }\Doi
  {10.1103/PhysRevA.91.022312} {\bibfield  {journal} {\bibinfo  {journal}
  {Phys. Rev. A},\ }\textbf {\bibinfo {volume} {91}},\ \bibinfo {pages}
  {022312} (\bibinfo {year} {2015})}\BibitemShut {NoStop}%
\bibitem [{\citenamefont {Gao}\ \emph {et~al.}(2012)\citenamefont {Gao},
  \citenamefont {Fallahi}, \citenamefont {Togan}, \citenamefont
  {Miguel-Sanchez},\ and\ \citenamefont {Imamoglu}}]{gao-nature-12}%
  \BibitemOpen
  \bibfield  {author} {\bibinfo {author} {\bibfnamefont {W.~B.}\ \bibnamefont
  {Gao}}, \bibinfo {author} {\bibfnamefont {P.}~\bibnamefont {Fallahi}},
  \bibinfo {author} {\bibfnamefont {E.}~\bibnamefont {Togan}}, \bibinfo
  {author} {\bibfnamefont {J.}~\bibnamefont {Miguel-Sanchez}}, \ and\ \bibinfo
  {author} {\bibfnamefont {A.}~\bibnamefont {Imamoglu}},\ }\href
  {http://dx.doi.org/10.1038/nature11573} {\bibfield  {journal} {\bibinfo
  {journal} {Nature},\ }\textbf {\bibinfo {volume} {491}},\ \bibinfo {pages}
  {426} (\bibinfo {year} {2012})}\BibitemShut {NoStop}%
\bibitem [{\citenamefont {Mandel}\ \emph {et~al.}(2003)\citenamefont {Mandel},
  \citenamefont {Greiner}, \citenamefont {Widera}, \citenamefont {Rom},
  \citenamefont {H{\"a}nsch},\ and\ \citenamefont {Bloch}}]{mandel-nature-03}%
  \BibitemOpen
  \bibfield  {author} {\bibinfo {author} {\bibfnamefont {O.}~\bibnamefont
  {Mandel}}, \bibinfo {author} {\bibfnamefont {M.}~\bibnamefont {Greiner}},
  \bibinfo {author} {\bibfnamefont {A.}~\bibnamefont {Widera}}, \bibinfo
  {author} {\bibfnamefont {T.}~\bibnamefont {Rom}}, \bibinfo {author}
  {\bibfnamefont {T.~W.}\ \bibnamefont {H{\"a}nsch}}, \ and\ \bibinfo {author}
  {\bibfnamefont {I.}~\bibnamefont {Bloch}},\ }\href
  {http://dx.doi.org/10.1038/nature02008} {\bibfield  {journal} {\bibinfo
  {journal} {Nature},\ }\textbf {\bibinfo {volume} {425}},\ \bibinfo {pages}
  {937} (\bibinfo {year} {2003})}\BibitemShut {NoStop}%
\bibitem [{\citenamefont {Laflamme}\ \emph {et~al.}(1998)\citenamefont
  {Laflamme}, \citenamefont {Knill}, \citenamefont {Zurek}, \citenamefont
  {Catasti},\ and\ \citenamefont {Mariappan}}]{laflamme-rsta-98}%
  \BibitemOpen
  \bibfield  {author} {\bibinfo {author} {\bibfnamefont {R.}~\bibnamefont
  {Laflamme}}, \bibinfo {author} {\bibfnamefont {E.}~\bibnamefont {Knill}},
  \bibinfo {author} {\bibfnamefont {W.~H.}\ \bibnamefont {Zurek}}, \bibinfo
  {author} {\bibfnamefont {P.}~\bibnamefont {Catasti}}, \ and\ \bibinfo
  {author} {\bibfnamefont {S.}~\bibnamefont {Mariappan}},\ }\Doi
  {10.1098/rsta.1998.0257} {\bibfield  {journal} {\bibinfo  {journal} {Philos.
  Trans. R. Soc. London, Ser A},\ }\textbf {\bibinfo {volume} {356}},\ \bibinfo
  {pages} {1941} (\bibinfo {year} {1998})}\BibitemShut {NoStop}%
\bibitem [{\citenamefont {Peng}\ \emph {et~al.}(2010)\citenamefont {Peng},
  \citenamefont {Zhang}, \citenamefont {Du},\ and\ \citenamefont
  {Suter}}]{peng-pra-10}%
  \BibitemOpen
  \bibfield  {author} {\bibinfo {author} {\bibfnamefont {X.}~\bibnamefont
  {Peng}}, \bibinfo {author} {\bibfnamefont {J.}~\bibnamefont {Zhang}},
  \bibinfo {author} {\bibfnamefont {J.}~\bibnamefont {Du}}, \ and\ \bibinfo
  {author} {\bibfnamefont {D.}~\bibnamefont {Suter}},\ }\Doi
  {10.1103/PhysRevA.81.042327} {\bibfield  {journal} {\bibinfo  {journal}
  {Phys. Rev. A},\ }\textbf {\bibinfo {volume} {81}},\ \bibinfo {pages}
  {042327} (\bibinfo {year} {2010})}\BibitemShut {NoStop}%
\bibitem [{\citenamefont {Rao}\ and\ \citenamefont
  {Kumar}(2012)}]{rao-ijqi-12}%
  \BibitemOpen
  \bibfield  {author} {\bibinfo {author} {\bibfnamefont {K.~R.~K.}\
  \bibnamefont {Rao}}\ and\ \bibinfo {author} {\bibfnamefont {A.}~\bibnamefont
  {Kumar}},\ }\Doi {10.1142/S0219749912500396} {\bibfield  {journal} {\bibinfo
  {journal} {Int. J. Quantum Info.},\ }\textbf {\bibinfo {volume} {10}},\
  \bibinfo {pages} {1250039} (\bibinfo {year} {2012})}\BibitemShut {NoStop}%
\bibitem [{\citenamefont {Das}\ \emph {et~al.}(2015)\citenamefont {Das},
  \citenamefont {Dogra}, \citenamefont {Dorai},\ and\ \citenamefont
  {Arvind}}]{dogra-pra-15-1}%
  \BibitemOpen
  \bibfield  {author} {\bibinfo {author} {\bibfnamefont {D.}~\bibnamefont
  {Das}}, \bibinfo {author} {\bibfnamefont {S.}~\bibnamefont {Dogra}}, \bibinfo
  {author} {\bibfnamefont {K.}~\bibnamefont {Dorai}}, \ and\ \bibinfo {author}
  {\bibnamefont {Arvind}},\ }\Doi {10.1103/PhysRevA.92.022307} {\bibfield
  {journal} {\bibinfo  {journal} {Phys. Rev. A},\ }\textbf {\bibinfo {volume}
  {92}},\ \bibinfo {pages} {022307} (\bibinfo {year} {2015})}\BibitemShut
  {NoStop}%
\bibitem [{\citenamefont {Xin}\ \emph {et~al.}(2018)\citenamefont {Xin},
  \citenamefont {Pedernales}, \citenamefont {Solano},\ and\ \citenamefont
  {Long}}]{xin-pra-18}%
  \BibitemOpen
  \bibfield  {author} {\bibinfo {author} {\bibfnamefont {T.}~\bibnamefont
  {Xin}}, \bibinfo {author} {\bibfnamefont {J.~S.}\ \bibnamefont {Pedernales}},
  \bibinfo {author} {\bibfnamefont {E.}~\bibnamefont {Solano}}, \ and\ \bibinfo
  {author} {\bibfnamefont {G.-L.}\ \bibnamefont {Long}},\ }\Doi
  {10.1103/PhysRevA.97.022322} {\bibfield  {journal} {\bibinfo  {journal}
  {Phys. Rev. A},\ }\textbf {\bibinfo {volume} {97}},\ \bibinfo {pages}
  {022322} (\bibinfo {year} {2018})}\BibitemShut {NoStop}%
\bibitem [{\citenamefont {Kampermann}\ \emph {et~al.}(2010)\citenamefont
  {Kampermann}, \citenamefont {Bru\ss{}}, \citenamefont {Peng},\ and\
  \citenamefont {Suter}}]{kampermann-pra-10}%
  \BibitemOpen
  \bibfield  {author} {\bibinfo {author} {\bibfnamefont {H.}~\bibnamefont
  {Kampermann}}, \bibinfo {author} {\bibfnamefont {D.}~\bibnamefont
  {Bru\ss{}}}, \bibinfo {author} {\bibfnamefont {X.}~\bibnamefont {Peng}}, \
  and\ \bibinfo {author} {\bibfnamefont {D.}~\bibnamefont {Suter}},\ }\Doi
  {10.1103/PhysRevA.81.040304} {\bibfield  {journal} {\bibinfo  {journal}
  {Phys. Rev. A},\ }\textbf {\bibinfo {volume} {81}},\ \bibinfo {pages}
  {040304} (\bibinfo {year} {2010})}\BibitemShut {NoStop}%
\bibitem [{\citenamefont {Bru\ss{}}(2002)}]{dagmar-jmp-02}%
  \BibitemOpen
  \bibfield  {author} {\bibinfo {author} {\bibfnamefont {D.}~\bibnamefont
  {Bru\ss{}}},\ }\Doi {10.1063/1.1494474} {\bibfield  {journal} {\bibinfo
  {journal} {J. Math. Phys.},\ }\textbf {\bibinfo {volume} {43}},\ \bibinfo
  {pages} {4237} (\bibinfo {year} {2002})}\BibitemShut {NoStop}%
\bibitem [{\citenamefont {Sackett}\ \emph {et~al.}(2000)\citenamefont
  {Sackett}, \citenamefont {Kielpinski}, \citenamefont {King}, \citenamefont
  {Langer}, \citenamefont {Meyer}, \citenamefont {Myatt}, \citenamefont {Rowe},
  \citenamefont {Turchette}, \citenamefont {Itano}, \citenamefont {Wineland},\
  and\ \citenamefont {Monroe}}]{sackett-nature-00}%
  \BibitemOpen
  \bibfield  {author} {\bibinfo {author} {\bibfnamefont {C.~A.}\ \bibnamefont
  {Sackett}}, \bibinfo {author} {\bibfnamefont {D.}~\bibnamefont {Kielpinski}},
  \bibinfo {author} {\bibfnamefont {B.~E.}\ \bibnamefont {King}}, \bibinfo
  {author} {\bibfnamefont {C.}~\bibnamefont {Langer}}, \bibinfo {author}
  {\bibfnamefont {V.}~\bibnamefont {Meyer}}, \bibinfo {author} {\bibfnamefont
  {C.~J.}\ \bibnamefont {Myatt}}, \bibinfo {author} {\bibfnamefont
  {M.}~\bibnamefont {Rowe}}, \bibinfo {author} {\bibfnamefont {Q.~A.}\
  \bibnamefont {Turchette}}, \bibinfo {author} {\bibfnamefont {W.~M.}\
  \bibnamefont {Itano}}, \bibinfo {author} {\bibfnamefont {D.~J.}\ \bibnamefont
  {Wineland}}, \ and\ \bibinfo {author} {\bibfnamefont {C.}~\bibnamefont
  {Monroe}},\ }\href {http://dx.doi.org/10.1038/35005011} {\bibfield  {journal}
  {\bibinfo  {journal} {Nature},\ }\textbf {\bibinfo {volume} {404}},\ \bibinfo
  {pages} {256} (\bibinfo {year} {2000})}\BibitemShut {NoStop}%
\bibitem [{\citenamefont {Wootters}(2001)}]{wootters-qic-01}%
  \BibitemOpen
  \bibfield  {author} {\bibinfo {author} {\bibfnamefont {W.~K.}\ \bibnamefont
  {Wootters}},\ }\href {http://dl.acm.org/citation.cfm?id=2011326.2011329}
  {\bibfield  {journal} {\bibinfo  {journal} {Quantum Info. Comput.},\ }\textbf
  {\bibinfo {volume} {1}},\ \bibinfo {pages} {27} (\bibinfo {year}
  {2001})}\BibitemShut {NoStop}%
\bibitem [{\citenamefont {Walborn}\ \emph {et~al.}(2006)\citenamefont
  {Walborn}, \citenamefont {Souto~Ribeiro}, \citenamefont {Davidovich},
  \citenamefont {Mintert},\ and\ \citenamefont
  {Buchleitner}}]{walborn-nature-06}%
  \BibitemOpen
  \bibfield  {author} {\bibinfo {author} {\bibfnamefont {S.~P.}\ \bibnamefont
  {Walborn}}, \bibinfo {author} {\bibfnamefont {P.~H.}\ \bibnamefont
  {Souto~Ribeiro}}, \bibinfo {author} {\bibfnamefont {L.}~\bibnamefont
  {Davidovich}}, \bibinfo {author} {\bibfnamefont {F.}~\bibnamefont {Mintert}},
  \ and\ \bibinfo {author} {\bibfnamefont {A.}~\bibnamefont {Buchleitner}},\
  }\href {http://dx.doi.org/10.1038/nature04627} {\bibfield  {journal}
  {\bibinfo  {journal} {Nature},\ }\textbf {\bibinfo {volume} {440}},\ \bibinfo
  {pages} {1022} (\bibinfo {year} {2006})}\BibitemShut {NoStop}%
\bibitem [{\citenamefont {Filgueiras}\ \emph {et~al.}(2012)\citenamefont
  {Filgueiras}, \citenamefont {Maciel}, \citenamefont {Auccaise}, \citenamefont
  {Vianna}, \citenamefont {Sarthour},\ and\ \citenamefont
  {Oliveira}}]{filgueiras-qip-12}%
  \BibitemOpen
  \bibfield  {author} {\bibinfo {author} {\bibfnamefont {J.~G.}\ \bibnamefont
  {Filgueiras}}, \bibinfo {author} {\bibfnamefont {T.~O.}\ \bibnamefont
  {Maciel}}, \bibinfo {author} {\bibfnamefont {R.~E.}\ \bibnamefont
  {Auccaise}}, \bibinfo {author} {\bibfnamefont {R.~O.}\ \bibnamefont
  {Vianna}}, \bibinfo {author} {\bibfnamefont {R.~S.}\ \bibnamefont
  {Sarthour}}, \ and\ \bibinfo {author} {\bibfnamefont {I.~S.}\ \bibnamefont
  {Oliveira}},\ }\Doi {10.1007/s11128-011-0341-z} {\bibfield  {journal}
  {\bibinfo  {journal} {Quant. Inf. Proc.},\ }\textbf {\bibinfo {volume}
  {11}},\ \bibinfo {pages} {1883} (\bibinfo {year} {2012})}\BibitemShut
  {NoStop}%
\bibitem [{\citenamefont {Bourennane}\ \emph {et~al.}(2004)\citenamefont
  {Bourennane}, \citenamefont {Eibl}, \citenamefont {Kurtsiefer}, \citenamefont
  {Gaertner}, \citenamefont {Weinfurter}, \citenamefont {G\"uhne},
  \citenamefont {Hyllus}, \citenamefont {Bru\ss{}}, \citenamefont
  {Lewenstein},\ and\ \citenamefont {Sanpera}}]{bourennane-prl-04}%
  \BibitemOpen
  \bibfield  {author} {\bibinfo {author} {\bibfnamefont {M.}~\bibnamefont
  {Bourennane}}, \bibinfo {author} {\bibfnamefont {M.}~\bibnamefont {Eibl}},
  \bibinfo {author} {\bibfnamefont {C.}~\bibnamefont {Kurtsiefer}}, \bibinfo
  {author} {\bibfnamefont {S.}~\bibnamefont {Gaertner}}, \bibinfo {author}
  {\bibfnamefont {H.}~\bibnamefont {Weinfurter}}, \bibinfo {author}
  {\bibfnamefont {O.}~\bibnamefont {G\"uhne}}, \bibinfo {author} {\bibfnamefont
  {P.}~\bibnamefont {Hyllus}}, \bibinfo {author} {\bibfnamefont
  {D.}~\bibnamefont {Bru\ss{}}}, \bibinfo {author} {\bibfnamefont
  {M.}~\bibnamefont {Lewenstein}}, \ and\ \bibinfo {author} {\bibfnamefont
  {A.}~\bibnamefont {Sanpera}},\ }\Doi {10.1103/PhysRevLett.92.087902}
  {\bibfield  {journal} {\bibinfo  {journal} {Phys. Rev. Lett.},\ }\textbf
  {\bibinfo {volume} {92}},\ \bibinfo {pages} {087902} (\bibinfo {year}
  {2004})}\BibitemShut {NoStop}%
\bibitem [{\citenamefont {D\"ur}\ and\ \citenamefont
  {Cirac}(2001)}]{dur-jpamg-01}%
  \BibitemOpen
  \bibfield  {author} {\bibinfo {author} {\bibfnamefont {W.}~\bibnamefont
  {D\"ur}}\ and\ \bibinfo {author} {\bibfnamefont {J.~I.}\ \bibnamefont
  {Cirac}},\ }\href {http://stacks.iop.org/0305-4470/34/i=35/a=310} {\bibfield
  {journal} {\bibinfo  {journal} {J. Phys. A: Math. Gen.},\ }\textbf {\bibinfo
  {volume} {34}},\ \bibinfo {pages} {6837} (\bibinfo {year}
  {2001})}\BibitemShut {NoStop}%
\bibitem [{\citenamefont {Altepeter}\ \emph {et~al.}(2005)\citenamefont
  {Altepeter}, \citenamefont {Jeffrey}, \citenamefont {Kwiat}, \citenamefont
  {Tanzilli}, \citenamefont {Gisin},\ and\ \citenamefont
  {Ac\'{\i}n}}]{altepeter-prl-05}%
  \BibitemOpen
  \bibfield  {author} {\bibinfo {author} {\bibfnamefont {J.~B.}\ \bibnamefont
  {Altepeter}}, \bibinfo {author} {\bibfnamefont {E.~R.}\ \bibnamefont
  {Jeffrey}}, \bibinfo {author} {\bibfnamefont {P.~G.}\ \bibnamefont {Kwiat}},
  \bibinfo {author} {\bibfnamefont {S.}~\bibnamefont {Tanzilli}}, \bibinfo
  {author} {\bibfnamefont {N.}~\bibnamefont {Gisin}}, \ and\ \bibinfo {author}
  {\bibfnamefont {A.}~\bibnamefont {Ac\'{\i}n}},\ }\Doi
  {10.1103/PhysRevLett.95.033601} {\bibfield  {journal} {\bibinfo  {journal}
  {Phys. Rev. Lett.},\ }\textbf {\bibinfo {volume} {95}},\ \bibinfo {pages}
  {033601} (\bibinfo {year} {2005})}\BibitemShut {NoStop}%
\bibitem [{\citenamefont {Spengler}\ \emph {et~al.}(2012)\citenamefont
  {Spengler}, \citenamefont {Huber}, \citenamefont {Brierley}, \citenamefont
  {Adaktylos},\ and\ \citenamefont {Hiesmayr}}]{spengler-pra-12}%
  \BibitemOpen
  \bibfield  {author} {\bibinfo {author} {\bibfnamefont {C.}~\bibnamefont
  {Spengler}}, \bibinfo {author} {\bibfnamefont {M.}~\bibnamefont {Huber}},
  \bibinfo {author} {\bibfnamefont {S.}~\bibnamefont {Brierley}}, \bibinfo
  {author} {\bibfnamefont {T.}~\bibnamefont {Adaktylos}}, \ and\ \bibinfo
  {author} {\bibfnamefont {B.~C.}\ \bibnamefont {Hiesmayr}},\ }\Doi
  {10.1103/PhysRevA.86.022311} {\bibfield  {journal} {\bibinfo  {journal}
  {Phys. Rev. A},\ }\textbf {\bibinfo {volume} {86}},\ \bibinfo {pages}
  {022311} (\bibinfo {year} {2012})}\BibitemShut {NoStop}%
\bibitem [{\citenamefont {Dai}\ \emph {et~al.}(2014)\citenamefont {Dai},
  \citenamefont {Len}, \citenamefont {Teo}, \citenamefont {Englert},\ and\
  \citenamefont {Krivitsky}}]{dai-prl-14}%
  \BibitemOpen
  \bibfield  {author} {\bibinfo {author} {\bibfnamefont {J.}~\bibnamefont
  {Dai}}, \bibinfo {author} {\bibfnamefont {Y.~L.}\ \bibnamefont {Len}},
  \bibinfo {author} {\bibfnamefont {Y.~S.}\ \bibnamefont {Teo}}, \bibinfo
  {author} {\bibfnamefont {B.-G.}\ \bibnamefont {Englert}}, \ and\ \bibinfo
  {author} {\bibfnamefont {L.~A.}\ \bibnamefont {Krivitsky}},\ }\Doi
  {10.1103/PhysRevLett.113.170402} {\bibfield  {journal} {\bibinfo  {journal}
  {Phys. Rev. Lett.},\ }\textbf {\bibinfo {volume} {113}},\ \bibinfo {pages}
  {170402} (\bibinfo {year} {2014})}\BibitemShut {NoStop}%
\bibitem [{\citenamefont {Chi}\ \emph {et~al.}(2010)\citenamefont {Chi},
  \citenamefont {Jeong}, \citenamefont {Kim}, \citenamefont {Lee},\ and\
  \citenamefont {Lee}}]{chi-pra-2010}%
  \BibitemOpen
  \bibfield  {author} {\bibinfo {author} {\bibfnamefont {D.~P.}\ \bibnamefont
  {Chi}}, \bibinfo {author} {\bibfnamefont {K.}~\bibnamefont {Jeong}}, \bibinfo
  {author} {\bibfnamefont {T.}~\bibnamefont {Kim}}, \bibinfo {author}
  {\bibfnamefont {K.}~\bibnamefont {Lee}}, \ and\ \bibinfo {author}
  {\bibfnamefont {S.}~\bibnamefont {Lee}},\ }\Doi {10.1103/PhysRevA.81.044302}
  {\bibfield  {journal} {\bibinfo  {journal} {Phys. Rev. A},\ }\textbf
  {\bibinfo {volume} {81}},\ \bibinfo {pages} {044302} (\bibinfo {year}
  {2010})}\BibitemShut {NoStop}%
\bibitem [{\citenamefont {Zhao}\ \emph {et~al.}(2013)\citenamefont {Zhao},
  \citenamefont {Zhang}, \citenamefont {Li-Jost},\ and\ \citenamefont
  {Fei}}]{zhao-pra-13}%
  \BibitemOpen
  \bibfield  {author} {\bibinfo {author} {\bibfnamefont {M.-J.}\ \bibnamefont
  {Zhao}}, \bibinfo {author} {\bibfnamefont {T.-G.}\ \bibnamefont {Zhang}},
  \bibinfo {author} {\bibfnamefont {X.}~\bibnamefont {Li-Jost}}, \ and\
  \bibinfo {author} {\bibfnamefont {S.-M.}\ \bibnamefont {Fei}},\ }\Doi
  {10.1103/PhysRevA.87.012316} {\bibfield  {journal} {\bibinfo  {journal}
  {Phys. Rev. A},\ }\textbf {\bibinfo {volume} {87}},\ \bibinfo {pages}
  {012316} (\bibinfo {year} {2013})}\BibitemShut {NoStop}%
\bibitem [{\citenamefont {Akbari-Kourbolagh}(2017)}]{akbari1}%
  \BibitemOpen
  \bibfield  {author} {\bibinfo {author} {\bibfnamefont {Y.}~\bibnamefont
  {Akbari-Kourbolagh}},\ }\Doi {10.1142/S0219749917500496} {\bibfield
  {journal} {\bibinfo  {journal} {Int. J. Quant. Inf.},\ }\textbf {\bibinfo
  {volume} {15}},\ \bibinfo {pages} {1750049} (\bibinfo {year}
  {2017})}\BibitemShut {NoStop}%
\bibitem [{\citenamefont {Akbari-Kourbolagh}\ and\ \citenamefont
  {Azhdargalam}(2018)}]{akbari2}%
  \BibitemOpen
  \bibfield  {author} {\bibinfo {author} {\bibfnamefont {Y.}~\bibnamefont
  {Akbari-Kourbolagh}}\ and\ \bibinfo {author} {\bibfnamefont {M.}~\bibnamefont
  {Azhdargalam}},\ }\Doi {10.1103/PhysRevA.97.042333} {\bibfield  {journal}
  {\bibinfo  {journal} {Phys. Rev. A},\ }\textbf {\bibinfo {volume} {97}},\
  \bibinfo {pages} {042333} (\bibinfo {year} {2018})}\BibitemShut {NoStop}%
\bibitem [{\citenamefont {D\"ur}\ \emph {et~al.}(2000)\citenamefont {D\"ur},
  \citenamefont {Vidal},\ and\ \citenamefont {Cirac}}]{dur-pra-00}%
  \BibitemOpen
  \bibfield  {author} {\bibinfo {author} {\bibfnamefont {W.}~\bibnamefont
  {D\"ur}}, \bibinfo {author} {\bibfnamefont {G.}~\bibnamefont {Vidal}}, \ and\
  \bibinfo {author} {\bibfnamefont {J.~I.}\ \bibnamefont {Cirac}},\ }\Doi
  {10.1103/PhysRevA.62.062314} {\bibfield  {journal} {\bibinfo  {journal}
  {Phys. Rev. A},\ }\textbf {\bibinfo {volume} {62}},\ \bibinfo {pages}
  {062314} (\bibinfo {year} {2000})}\BibitemShut {NoStop}%
\bibitem [{\citenamefont {Bennett}\ \emph {et~al.}(2000)\citenamefont
  {Bennett}, \citenamefont {Popescu}, \citenamefont {Rohrlich}, \citenamefont
  {Smolin},\ and\ \citenamefont {Thapliyal}}]{bennett-pra-00}%
  \BibitemOpen
  \bibfield  {author} {\bibinfo {author} {\bibfnamefont {C.~H.}\ \bibnamefont
  {Bennett}}, \bibinfo {author} {\bibfnamefont {S.}~\bibnamefont {Popescu}},
  \bibinfo {author} {\bibfnamefont {D.}~\bibnamefont {Rohrlich}}, \bibinfo
  {author} {\bibfnamefont {J.~A.}\ \bibnamefont {Smolin}}, \ and\ \bibinfo
  {author} {\bibfnamefont {A.~V.}\ \bibnamefont {Thapliyal}},\ }\Doi
  {10.1103/PhysRevA.63.012307} {\bibfield  {journal} {\bibinfo  {journal}
  {Phys. Rev. A},\ }\textbf {\bibinfo {volume} {63}},\ \bibinfo {pages}
  {012307} (\bibinfo {year} {2000})}\BibitemShut {NoStop}%
\bibitem [{\citenamefont {Wootters}(1998)}]{wootters-prl-98}%
  \BibitemOpen
  \bibfield  {author} {\bibinfo {author} {\bibfnamefont {W.~K.}\ \bibnamefont
  {Wootters}},\ }\Doi {10.1103/PhysRevLett.80.2245} {\bibfield  {journal}
  {\bibinfo  {journal} {Phys. Rev. Lett.},\ }\textbf {\bibinfo {volume} {80}},\
  \bibinfo {pages} {2245} (\bibinfo {year} {1998})}\BibitemShut {NoStop}%
\bibitem [{\citenamefont {Adhikari}\ \emph {et~al.}(2017)\citenamefont
  {Adhikari}, \citenamefont {Datta}, \citenamefont {Das},\ and\ \citenamefont
  {Agrawal}}]{adhikari-arxiv-17}%
  \BibitemOpen
  \bibfield  {author} {\bibinfo {author} {\bibfnamefont {S.}~\bibnamefont
  {Adhikari}}, \bibinfo {author} {\bibfnamefont {C.}~\bibnamefont {Datta}},
  \bibinfo {author} {\bibfnamefont {A.}~\bibnamefont {Das}}, \ and\ \bibinfo
  {author} {\bibfnamefont {P.}~\bibnamefont {Agrawal}},\ }\href@noop {}
  {\bibfield  {journal} {\bibinfo  {journal} {arXiv}} (\bibinfo {year}
  {2017})},\ \Eprint {http://arxiv.org/abs/1705.01377} {1705.01377}
  \BibitemShut {NoStop}%
\bibitem [{\citenamefont {Singh}\ \emph {et~al.}(2018)\citenamefont {Singh},
  \citenamefont {Singh}, \citenamefont {Dorai},\ and\ \citenamefont
  {Arvind}}]{singh-arxiv-18}%
  \BibitemOpen
  \bibfield  {author} {\bibinfo {author} {\bibfnamefont {A.}~\bibnamefont
  {Singh}}, \bibinfo {author} {\bibfnamefont {H.}~\bibnamefont {Singh}},
  \bibinfo {author} {\bibfnamefont {K.}~\bibnamefont {Dorai}}, \ and\ \bibinfo
  {author} {\bibnamefont {Arvind}},\ }\href {https://arxiv.org/abs/1804.09320}
  {\bibfield  {journal} {\bibinfo  {journal} {arXiv}} (\bibinfo {year}
  {2018})},\ \Eprint {http://arxiv.org/abs/1804.09320} {1804.09320}
  \BibitemShut {NoStop}%
\bibitem [{\citenamefont {Singh}\ \emph
  {et~al.}(2016){\natexlab{a}}\citenamefont {Singh}, \citenamefont {Arvind},\
  and\ \citenamefont {Dorai}}]{singh-pra-16}%
  \BibitemOpen
  \bibfield  {author} {\bibinfo {author} {\bibfnamefont {A.}~\bibnamefont
  {Singh}}, \bibinfo {author} {\bibnamefont {Arvind}}, \ and\ \bibinfo {author}
  {\bibfnamefont {K.}~\bibnamefont {Dorai}},\ }\Doi
  {10.1103/PhysRevA.94.062309} {\bibfield  {journal} {\bibinfo  {journal}
  {Phys. Rev. A},\ }\textbf {\bibinfo {volume} {94}},\ \bibinfo {pages}
  {062309} (\bibinfo {year} {2016}{\natexlab{a}})}\BibitemShut {NoStop}%
\bibitem [{\citenamefont {Leskowitz}\ and\ \citenamefont
  {Mueller}(2004)}]{leskowitz-pra-04}%
  \BibitemOpen
  \bibfield  {author} {\bibinfo {author} {\bibfnamefont {G.~M.}\ \bibnamefont
  {Leskowitz}}\ and\ \bibinfo {author} {\bibfnamefont {L.~J.}\ \bibnamefont
  {Mueller}},\ }\Doi {10.1103/PhysRevA.69.052302} {\bibfield  {journal}
  {\bibinfo  {journal} {Phys. Rev. A},\ }\textbf {\bibinfo {volume} {69}},\
  \bibinfo {pages} {052302} (\bibinfo {year} {2004})}\BibitemShut {NoStop}%
\bibitem [{\citenamefont {Weinstein}(2010)}]{weinstein-pra-10}%
  \BibitemOpen
  \bibfield  {author} {\bibinfo {author} {\bibfnamefont {Y.~S.}\ \bibnamefont
  {Weinstein}},\ }\Doi {10.1103/PhysRevA.82.032326} {\bibfield  {journal}
  {\bibinfo  {journal} {Phys. Rev. A},\ }\textbf {\bibinfo {volume} {82}},\
  \bibinfo {pages} {032326} (\bibinfo {year} {2010})}\BibitemShut {NoStop}%
\bibitem [{\citenamefont {Vidal}\ and\ \citenamefont
  {Werner}(2002)}]{vidal-pra-02}%
  \BibitemOpen
  \bibfield  {author} {\bibinfo {author} {\bibfnamefont {G.}~\bibnamefont
  {Vidal}}\ and\ \bibinfo {author} {\bibfnamefont {R.~F.}\ \bibnamefont
  {Werner}},\ }\Doi {10.1103/PhysRevA.65.032314} {\bibfield  {journal}
  {\bibinfo  {journal} {Phys. Rev. A},\ }\textbf {\bibinfo {volume} {65}},\
  \bibinfo {pages} {032314} (\bibinfo {year} {2002})}\BibitemShut {NoStop}%
\bibitem [{\citenamefont {Rungta}\ \emph {et~al.}(2001)\citenamefont {Rungta},
  \citenamefont {Bu\ifmmode~\check{z}\else \v{z}\fi{}ek}, \citenamefont
  {Caves}, \citenamefont {Hillery},\ and\ \citenamefont
  {Milburn}}]{rungta-pra-01}%
  \BibitemOpen
  \bibfield  {author} {\bibinfo {author} {\bibfnamefont {P.}~\bibnamefont
  {Rungta}}, \bibinfo {author} {\bibfnamefont {V.}~\bibnamefont
  {Bu\ifmmode~\check{z}\else \v{z}\fi{}ek}}, \bibinfo {author} {\bibfnamefont
  {C.~M.}\ \bibnamefont {Caves}}, \bibinfo {author} {\bibfnamefont
  {M.}~\bibnamefont {Hillery}}, \ and\ \bibinfo {author} {\bibfnamefont
  {G.~J.}\ \bibnamefont {Milburn}},\ }\Doi {10.1103/PhysRevA.64.042315}
  {\bibfield  {journal} {\bibinfo  {journal} {Phys. Rev. A},\ }\textbf
  {\bibinfo {volume} {64}},\ \bibinfo {pages} {042315} (\bibinfo {year}
  {2001})}\BibitemShut {NoStop}%
\bibitem [{\citenamefont {Ac\'{\i}n}\ \emph {et~al.}(2001)\citenamefont
  {Ac\'{\i}n}, \citenamefont {Bru\ss{}}, \citenamefont {Lewenstein},\ and\
  \citenamefont {Sanpera}}]{acin-prl-01}%
  \BibitemOpen
  \bibfield  {author} {\bibinfo {author} {\bibfnamefont {A.}~\bibnamefont
  {Ac\'{\i}n}}, \bibinfo {author} {\bibfnamefont {D.}~\bibnamefont {Bru\ss{}}},
  \bibinfo {author} {\bibfnamefont {M.}~\bibnamefont {Lewenstein}}, \ and\
  \bibinfo {author} {\bibfnamefont {A.}~\bibnamefont {Sanpera}},\ }\Doi
  {10.1103/PhysRevLett.87.040401} {\bibfield  {journal} {\bibinfo  {journal}
  {Phys. Rev. Lett.},\ }\textbf {\bibinfo {volume} {87}},\ \bibinfo {pages}
  {040401} (\bibinfo {year} {2001})}\BibitemShut {NoStop}%
\bibitem [{\citenamefont {Wong}\ and\ \citenamefont
  {Christensen}(2001)}]{wong-pra-01}%
  \BibitemOpen
  \bibfield  {author} {\bibinfo {author} {\bibfnamefont {A.}~\bibnamefont
  {Wong}}\ and\ \bibinfo {author} {\bibfnamefont {N.}~\bibnamefont
  {Christensen}},\ }\Doi {10.1103/PhysRevA.63.044301} {\bibfield  {journal}
  {\bibinfo  {journal} {Phys. Rev. A},\ }\textbf {\bibinfo {volume} {63}},\
  \bibinfo {pages} {044301} (\bibinfo {year} {2001})}\BibitemShut {NoStop}%
\bibitem [{\citenamefont {Li}(2012)}]{li-qip-12}%
  \BibitemOpen
  \bibfield  {author} {\bibinfo {author} {\bibfnamefont {D.}~\bibnamefont
  {Li}},\ }\Doi {10.1007/s11128-011-0256-8} {\bibfield  {journal} {\bibinfo
  {journal} {Quant. Inf. Proc.},\ }\textbf {\bibinfo {volume} {11}},\ \bibinfo
  {pages} {481} (\bibinfo {year} {2012})},\ ISSN \bibinfo {issn}
  {1573-1332}\BibitemShut {NoStop}%
\bibitem [{\citenamefont {Nielsen}\ and\ \citenamefont
  {Chuang}(2000)}]{nielsen-book-02}%
  \BibitemOpen
  \bibfield  {author} {\bibinfo {author} {\bibfnamefont {M.~A.}\ \bibnamefont
  {Nielsen}}\ and\ \bibinfo {author} {\bibfnamefont {I.~L.}\ \bibnamefont
  {Chuang}},\ }\href
  {https://www.cambridge.org/core/books/quantum-computation-and-quantum-information/01E10196D0A682A6AEFFEA52D53BE9AE}
  {\emph {\bibinfo {title} {Quantum Computation and Quantum Information}}}\
  (\bibinfo  {publisher} {Cambridge University Press},\ \bibinfo {year}
  {2000})\ ISBN \bibinfo {isbn} {0511976666}\BibitemShut {NoStop}%
\bibitem [{\citenamefont {Oliveira}\ \emph {et~al.}(2007)\citenamefont
  {Oliveira}, \citenamefont {Bonagamba}, \citenamefont {Sarthour},
  \citenamefont {Freitas},\ and\ \citenamefont {deAzevedo}}]{oliveira-book-07}%
  \BibitemOpen
  \bibfield  {author} {\bibinfo {author} {\bibfnamefont {I.~S.}\ \bibnamefont
  {Oliveira}}, \bibinfo {author} {\bibfnamefont {T.~J.}\ \bibnamefont
  {Bonagamba}}, \bibinfo {author} {\bibfnamefont {R.~S.}\ \bibnamefont
  {Sarthour}}, \bibinfo {author} {\bibfnamefont {J.~C.~C.}\ \bibnamefont
  {Freitas}}, \ and\ \bibinfo {author} {\bibfnamefont {E.~R.}\ \bibnamefont
  {deAzevedo}},\ }\href
  {https://www.sciencedirect.com/science/book/9780444527820} {\emph {\bibinfo
  {title} {NMR Quantum Information Processing}}}\ (\bibinfo  {publisher}
  {Elsevier},\ \bibinfo {address} {Linacre House, Jordan Hill, Oxford OX2 8DP,
  UK},\ \bibinfo {year} {2007})\BibitemShut {NoStop}%
\bibitem [{\citenamefont {Ernst}\ \emph {et~al.}(1990)\citenamefont {Ernst},
  \citenamefont {Bodenhausen},\ and\ \citenamefont {Wokaun}}]{ernst-book-90}%
  \BibitemOpen
  \bibfield  {author} {\bibinfo {author} {\bibfnamefont {R.~R.}\ \bibnamefont
  {Ernst}}, \bibinfo {author} {\bibfnamefont {G.}~\bibnamefont {Bodenhausen}},
  \ and\ \bibinfo {author} {\bibfnamefont {A.}~\bibnamefont {Wokaun}},\ }\href
  {https://www.amazon.com/Principles-Resonance-Dimensions-International-Monographs/dp/0198556470?SubscriptionId=0JYN1NVW651KCA56C102&tag=techkie-20&linkCode=xm2&camp=2025&creative=165953&creativeASIN=0198556470}
  {\emph {\bibinfo {title} {Principles of NMR in One and Two Dimensions}}}\
  (\bibinfo  {publisher} {Clarendon Press},\ \bibinfo {year} {1990})\ ISBN
  \bibinfo {isbn} {0198556470}\BibitemShut {NoStop}%
\bibitem [{\citenamefont {Cory}\ \emph {et~al.}(1998)\citenamefont {Cory},
  \citenamefont {Price},\ and\ \citenamefont {Havel}}]{cory-physD-98}%
  \BibitemOpen
  \bibfield  {author} {\bibinfo {author} {\bibfnamefont {D.~G.}\ \bibnamefont
  {Cory}}, \bibinfo {author} {\bibfnamefont {M.~D.}\ \bibnamefont {Price}}, \
  and\ \bibinfo {author} {\bibfnamefont {T.~F.}\ \bibnamefont {Havel}},\ }\Doi
  {10.1016/S0167-2789(98)00046-3} {\bibfield  {journal} {\bibinfo  {journal}
  {Physica D: Nonlinear Phenomena},\ }\textbf {\bibinfo {volume} {120}},\
  \bibinfo {pages} {82 } (\bibinfo {year} {1998})}\BibitemShut {NoStop}%
\bibitem [{\citenamefont {Mitra}\ \emph {et~al.}(2007)\citenamefont {Mitra},
  \citenamefont {Sivapriya},\ and\ \citenamefont {Kumar}}]{mitra-jmr-07}%
  \BibitemOpen
  \bibfield  {author} {\bibinfo {author} {\bibfnamefont {A.}~\bibnamefont
  {Mitra}}, \bibinfo {author} {\bibfnamefont {K.}~\bibnamefont {Sivapriya}}, \
  and\ \bibinfo {author} {\bibfnamefont {A.}~\bibnamefont {Kumar}},\ }\Doi
  {10.1016/j.jmr.2007.05.013} {\bibfield  {journal} {\bibinfo  {journal} {J.
  Magn. Reson.},\ }\textbf {\bibinfo {volume} {187}},\ \bibinfo {pages} {306 }
  (\bibinfo {year} {2007})}\BibitemShut {NoStop}%
\bibitem [{\citenamefont {Uhlmann}(1976)}]{uhlmann-rpmp-76}%
  \BibitemOpen
  \bibfield  {author} {\bibinfo {author} {\bibfnamefont {A.}~\bibnamefont
  {Uhlmann}},\ }\Doi {10.1016/0034-4877(76)90060-4} {\bibfield  {journal}
  {\bibinfo  {journal} {Rep. Math. Phys.},\ }\textbf {\bibinfo {volume} {9}},\
  \bibinfo {pages} {273 } (\bibinfo {year} {1976})}\BibitemShut {NoStop}%
\bibitem [{\citenamefont {Jozsa}(1994)}]{jozsa-jmo-94}%
  \BibitemOpen
  \bibfield  {author} {\bibinfo {author} {\bibfnamefont {R.}~\bibnamefont
  {Jozsa}},\ }\Doi {10.1080/09500349414552171} {\bibfield  {journal} {\bibinfo
  {journal} {J. Mod. Optics},\ }\textbf {\bibinfo {volume} {41}},\ \bibinfo
  {pages} {2315} (\bibinfo {year} {1994})}\BibitemShut {NoStop}%
\bibitem [{\citenamefont {Singh}\ \emph
  {et~al.}(2016){\natexlab{b}}\citenamefont {Singh}, \citenamefont {Arvind},\
  and\ \citenamefont {Dorai}}]{singh-pla-16}%
  \BibitemOpen
  \bibfield  {author} {\bibinfo {author} {\bibfnamefont {H.}~\bibnamefont
  {Singh}}, \bibinfo {author} {\bibnamefont {Arvind}}, \ and\ \bibinfo {author}
  {\bibfnamefont {K.}~\bibnamefont {Dorai}},\ }\Doi
  {10.1016/j.physleta.2016.07.046} {\bibfield  {journal} {\bibinfo  {journal}
  {Physics Letters A},\ }\textbf {\bibinfo {volume} {380}},\ \bibinfo {pages}
  {3051 } (\bibinfo {year} {2016}{\natexlab{b}})}\BibitemShut {NoStop}%
\end{thebibliography}

%

\end{document}